\documentclass[10pt,prl,aps,twocolumn,superscriptaddress,preprintnumbers]{revtex4-1}

\usepackage[utf8]{inputenc}
\usepackage{epigraph}
\usepackage{amsmath}
\usepackage{wrapfig}
\usepackage{epsfig}
\usepackage{amssymb}
\usepackage{graphicx}
\usepackage{slashed}
\usepackage{xcolor}
\usepackage{comment}
\usepackage{natbib}
\usepackage{enumitem}
\usepackage{tikz-cd,tikz}
\usepackage{float}
\usepackage{array,adjustbox,booktabs}
\usepackage{subcaption}
\usepackage{amsfonts}
\usepackage{physics}
\usepackage{listings}
\usepackage{calc}
\usepackage{soul}

\newcommand\beqa{\begin{eqnarray}}
\newcommand\eeqa{\end{eqnarray}}
\newcommand{\beq}{\begin{eqnarray}}
\newcommand{\eeq}{\end{eqnarray}}
\newcommand{\la}[1]{\label{#1}}
\newcommand{\eq}[1]{(\ref{#1})}

\newcommand{\res}[2]{\underset{#2}{\operatorname{res}}\left(#1\right)}

\newcommand\mathstyle{\lstset{
language=Mathematica,
basicstyle={\scriptsize\def\fvm@Scale{.5}\fontfamily{fvm}\selectfont},
otherkeywords={self},
keywordstyle=\ttb\scriptsize\color{deepblue},
emph={MyClass,__init__},
emphstyle=\ttb\color{deepred},
backgroundcolor=\color{pink!20!white},
stringstyle=\color{deepgreen},
commentstyle=\color{SkyBlue3!70!PaleGreen4},
frame=tb,
showstringspaces=false
}}

\newwrite\todofile
\immediate\openout\todofile=\jobname.tdo

\newcounter{todocounter}

\newcommand{\printtodos}{
        \section*{To-Do List}
        \immediate\closeout\todofile
        \input{\jobname.tdo}
}
\lstnewenvironment{mathematica}[1][]
{
\mathstyle
\lstset{#1}
}
{}

\newcolumntype{C}[1]{>{\centering\arraybackslash}m{#1}}

\newcommand{\algsl}{\mathfrak{sl}}
\newcommand{\ii}{i}

\newcommand{\baxterQ}{\mathbb{Q}}

\newcommand{\fQ}{\mathcal{Q}}
\newcommand{\bQ}{\mathbf{Q}}
\newcommand{\bP}{\mathbf{P}}

\newcommand{\betheQ}{\mathbb{Q}}

\usepackage[hidelinks,colorlinks=true]{hyperref}

\begin{document}
\title{Long Range Asymptotic Baxter-Bethe Ansatz for $\mathcal{N}=4$ BFKL}

\author{Simon Ekhammar}
\email{simon.ekhammar@kcl.ac.uk}
\affiliation{Department of Mathematics, King’s College London
Strand WC2R 2LS, London, UK}
\author{Nikolay Gromov}
\email{nikolay.gromov@kcl.ac.uk}
\affiliation{Department of Mathematics, King’s College London
Strand WC2R 2LS, London, UK}
\author{Michelangelo Preti}
 \email{michelangelo.preti@stonybrook.edu}
\affiliation{Dipartimento di Fisica, Universit\`a di Torino and INFN - Sezione di Torino, Via P. Giuria 1, Torino 10125, Italy}
\affiliation{C. N. Yang Institute for Theoretical Physics, Stony Brook University, Stony Brook, New York 11794, USA}
\affiliation{Simons Center for Geometry and Physics, Stony Brook University, Stony Brook, New York 11794, USA
}

\begin{abstract}
We demonstrate that the Balitsky-Fadin-Kuraev-Lipatov regime of maximally supersymmetric Yang-Mills theory can be explicitly solved up to the $L+1$ order in weak coupling by uncovering a novel long-range asymptotic Baxter-Bethe ansatz for trajectories with $L$ scalar fields. The set of equations we have found is reminiscent of the Beisert-Eden-Staudacher equations for local operators but instead applies to non-local operators corresponding to the horizontal Regge trajectories.
We also verify and give new predictions for the light-ray operator spectrum by resummation of the leading singularities in our result. 
\end{abstract}


\maketitle

\section{Introduction}

Regge trajectories play a crucial role in modern non-perturbative physics. In hadron spectroscopy, they relate spin and mass squared aiding in the discovery of various mesons and baryons.
At the same time, they are also key to understanding high-energy scattering amplitudes. However, their theoretical study is challenging due to the need for analytic continuation in spin and for them being intrinsically non-perturbative.

The Balitsky–Fadin–Kuraev–Lipatov (BFKL) equation offers a powerful resummation technique to address some of these issues in quantum chromodynamics (QCD) \cite{kuraev1976multiregge,kuraev1977pomeranchuk,balitsky1978pomeranchuk}. In some cases, it allows accurate predictions of cross-sections in processes such as deep inelastic scattering and forward jet production, even though it is worked out only to the leading (LO) and next-to-leading orders in QCD \cite{Jaroszewicz:1982gr,Lipatov:1985uk,Fadin:1998py,Ciafaloni:1998gs,Kotikov:2000pm,Kotikov:2002ab}. 

Gauge theories that allow fully non-perturbative integrability methods are ideal laboratories for exploring Regge trajectories and resummation techniques. One of the most prominent example is planar
${\cal N}=4$ Super Yang-Mills (SYM). Due to integrability many observables in $\mathcal{N}=4$ SYM are solvable to all orders in the 't Hooft coupling $\lambda$, at the same time this theory also shares many features with QCD. An important example is the Kotikov-Lipatov maximal transcendentality principle \cite{Kotikov:2002ab} stating that the results of ${\cal N}=4$ frequently give direct prediction for their counterpart in QCD for the most complicated, maximally transcendental part \cite{Kotikov:2001sc,Kotikov:2010nd}.
In particular, integrability was used to obtain analytic expressions for eigenvalues of the NNLO BFKL~\cite{Gromov:2015vua}
equation in ${\cal N}=4$ SYM, confirmed independently in~\cite{Caron-Huot:2016tzz}\footnote{see also~\cite{Velizhanin:2015xsa}.}, providing precious insights into QCD.
Furthermore, integrability in $\mathcal{N}=4$ SYM enables a completely non-perturbative study of Regge trajectories \cite{Alfimov:2014bwa,Gromov:2015wca,gromov2015pomeron,Alfimov:2018cms,Klabbers:2023zdz,Homrich:2022mmd,Henriksson:2023cnh}.

The integrability is tightly related to the conformal symmetry of the theory. In general conformal field theories (CFTs) it is possible to study
Regge physics through analytic continuation of the scaling dimensions $\Delta$ in spin \cite{Cornalba:2007fs,Costa:2012cb}. 
To perform this analytic continuation, 
as argued in~\cite{Balitsky:1987bk,kravchuk2018light}, one should study non-local light-ray operators, which give direct access to non-integer spins in perturbation theory. Light-ray operators can be constructed as null Wilson lines \cite{kravchuk2018light,balitsky2023two,Kologlu:2019mfz,Chang:2020qpj,caron2023detectors} and allows one to closely approach the BFKL regime. However, as we discuss in this letter, in general there are degeneracies at weak coupling, creating additional challenges~\cite{caron2015does,caron2023detectors,Korchemsky:2003rc,Klabbers:2023zdz}.

\paragraph{Integrability and Pomeron.}
Integrability in $4$D gauge theories was first observed in the BFKL regime by Lipatov \cite{Lipatov:1993yb} and further explored by \cite{Faddeev:1994zg,Korchemsky_1995}. A key finding was the Baxter equation
\begin{equation}\label{eq:LipatovBaxter}
4u^2 \left(q^{++}+q^{--}\right)=\left(8u^2-{\Delta^2+1}{}\right)q
\end{equation}
where we introduced shift of argument notation $f^\pm = f(u\pm\tfrac{i}{2})$ and
$f^{\pm\pm} = f(u\pm i)$ or more generally $f^{[n]}=f(u+i n/2)$. 
Solution of \eq{eq:LipatovBaxter} gives the famous LO Pomeron eigenvalue
via~\footnote{for a specific solution $q(u)=2 i u \, _3F_2\left(i u+1,\frac{1}{2}-\frac{\Delta }{2},\frac{\Delta
   }{2}+\frac{1}{2};1,2;1\right)$}
\begin{equation}\label{pomeroneigenvalue}
\left.-i\partial_u\log u q^{--}\right|_{u\to 0}=\psi\left( \tfrac{1-\Delta}{2}  \right) + \psi\left( \tfrac{1+\Delta}{2} \right) + 2\gamma \equiv \chi\,,
\end{equation}
where $\psi$ is Digamma and $\gamma$ is the Euler constant. 
Notably, the above equations are the same for QCD and ${\cal N}=4$ SYM~\cite{Bartels:2009vkz}. In this paper we present the multi-loop generalization of these equation for parity symmetric states, containing $L$ scalar fields. This is analogous to how the Asymptotic Bethe Ansatz (ABA) \cite{Beisert:2005fw}, or Beisert-Eden-Staudacher equation \cite{Beisert:2006ez}, generalizes the rational spin chain of Minahan and Zarembo \cite{Minahan:2002ve} to higher loops. In particular, for the top horizontal trajectory i.e. for $\omega = S+L-1\to 0$ we find 
\begin{equation}\label{demoeq}
\omega =\!\! 
\sum_{k=1,2}\!\!\left(\!4g \cos\!\frac{\pi  n_k}{L+2} -16  g^2\chi(\Delta)\frac{\sin ^2\!\frac{\pi  n_k}{L+2}}{L+2}\!\right)\!+{\cal O}(g^3)
\end{equation}
with $g=\frac{\sqrt{\lambda}}{4 \pi}$ the 't Hooft coupling and $1\leq n_1<n_2 \leq L+1$ with $n_1+n_2\in 2\,\mathbb{Z}$. This reproduces \eqref{pomeroneigenvalue} for $n_1=1,n_2=3$ and $L=2$. Notice that the leading term is linear in $g$ and independent on $\Delta$ as first noticed for the $L=3$ case in \cite{Klabbers:2023zdz}. In this letter, we describe how to explicitly compute all terms up to $g^{L+1}$ for general $L$.

\paragraph{Regge trajectories.}
For simplicity in this letter we focus on the generalization of the BFKL regime for operators in $\mathfrak{sl}(2)$ sector in which the local operators take the schematic form ${\rm Tr} D_+^S Z^L+{\rm\it perm}$ and furthermore are parity symmetric. We reserve further generalizations beyond parity symmetric states, as in \cite{Derkachov:2002wz}, and outside $\mathfrak{sl}(2)$-sector for the future \cite{UpcommingLong}.
\begin{figure}[!t]
    \centering
    \includegraphics[width=\columnwidth]{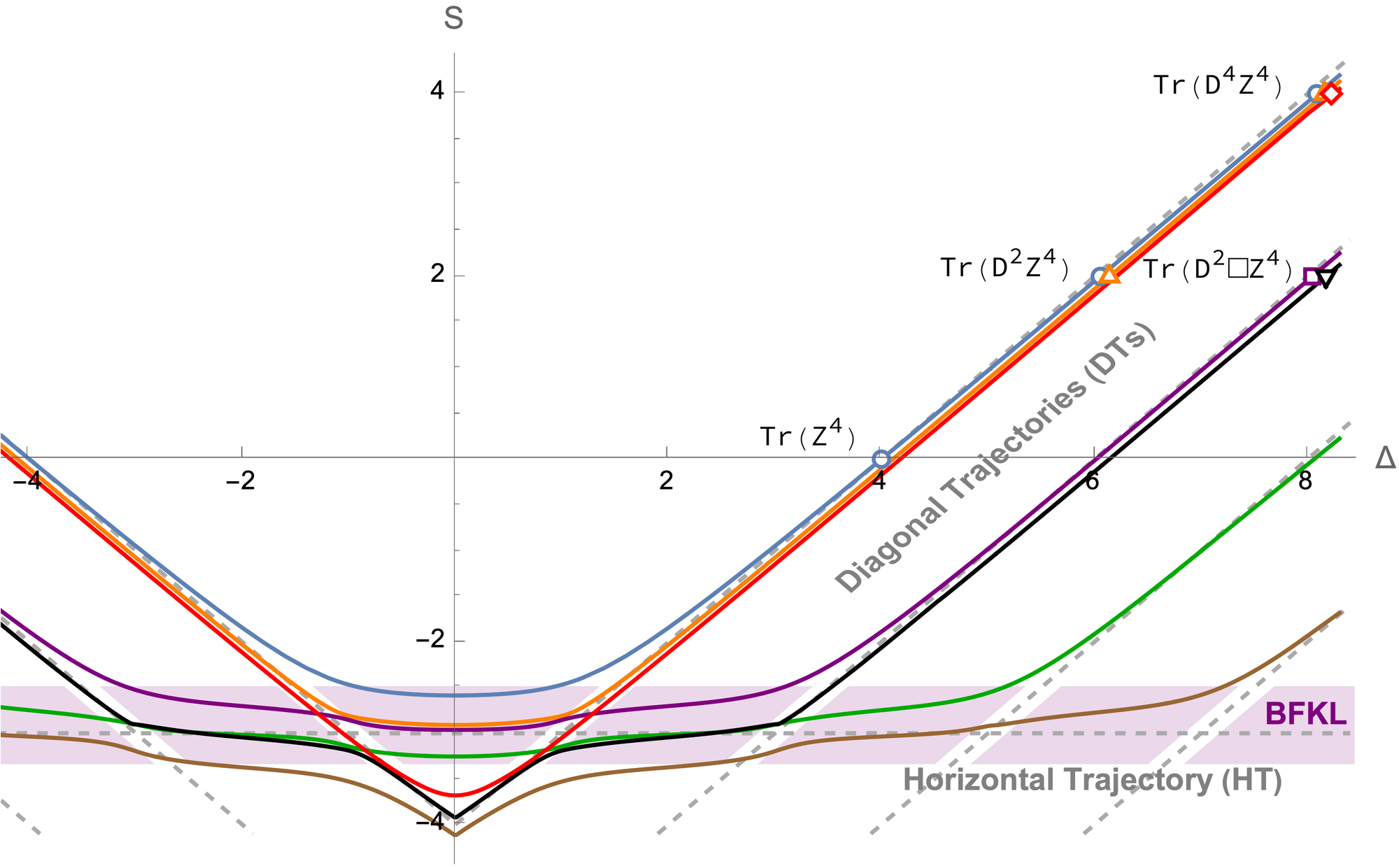}
    \caption{The typical Regge trajectory (using $L=4$ numerics for demonstration). The diagonal parts (DGLAP or DT) of trajectories contain local operators. The first horizontal part of the trajectories (BFKL or HT), is where this letter is focused on.}
    \label{fig:enter-label1}
\end{figure}
In perturbation theory these operators have dimension  $\Delta=S+L+\gamma(S)$.
To reach the BFKL regime one has to analytically continue $\Delta$ in spin $S$ and consider the inverse function $S(\Delta)$. The function $S(\Delta)$ is a multi-valued function, whose Riemann surface is invariant under $\Delta\to-\Delta$ (see FIG.~\ref{fig:enter-label1} for $L=4$). For general $L$ the number of local operators increases with $S$ and there are infinitely many diagonal trajectories (DTs), we refer to this as the Dokshitzer–Gribov–Lipatov–Altarelli–Parisi (DGLAP) regime. However, only a finite number of them branch to horizontal trajectories (HTs) at $S=-L+1$, and those are the main focus of this paper. It is important to notice that not only the $\mathfrak{sl}(2)$ local operators appear at special points of this Riemann surface. One can also find operators of ``higher twist", which are roughly of the form ${\rm Tr} \Box^m D_+^S Z^L$ (see e.g.~\cite{Gromov:2015wca,Klabbers:2023zdz}).

\section{Integrability of ${\cal N}=4$ SYM}
Unlike in QCD, integrability of ${\cal N}=4$ persists to arbitrary $g$ in the planar limit. There has been many integrability based techniques developed to address BFKL physics (see~\cite{Gromov:2017blm} for a review), we will use the truly non-perturbative Quantum Spectral Curve (QSC)~\cite{Gromov:2013pga,Gromov:2014caa} for our exploration.

Before the QSC, integrability in the BFKL regime \eqref{eq:LipatovBaxter} and for local operators~\cite{Minahan:2002ve} co-existed almost independently, even though the leading singularities in both regimes where known to be related \cite{Kotikov:2007cy}. The reason being that from the local operators perspective the BFKL regime is completely non-perturbative and requires re-summation of infinitely many terms of the form $g^{2n}/\omega^n$ (for $L=2$) where $\omega=S+L-1$, whereas the ABA gives a powerful and concise description of the spectrum only up to so-called wrapping order $\sim g^{2L+4}$. 

\begin{figure}[t]
    \centering
    \includegraphics[scale=0.92]{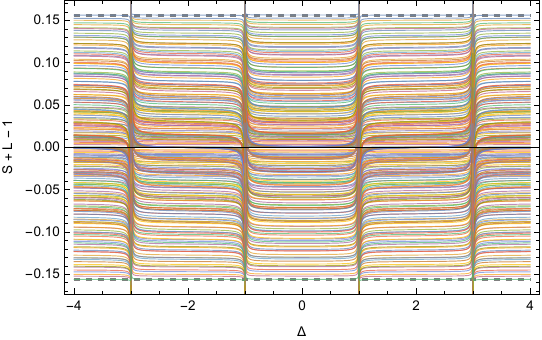}
    \caption{Spectrum of $225$ states for $L=30$ at $g=0.02$. At weak coupling they spread between the bounds $\pm 4 g \left(\cos \frac{\pi }{L+2}+\cos \frac{3 \pi }{L+2}\right)$ (dashed gray lines). For the plot we use only first $3$ terms in small $g$, but in principle up to $g^{31}$ can be computed analytically with our method.}
    \label{fig:enter-label}
\end{figure}

With the QSC, valid at any $g$, one can numerically interpolate between the DGLAP (DTs) and BFKL (HTs) branches of the same trajectory (see FIG.~\ref{fig:enter-label}). However, analytic results have so far been out of reach for general operators. In the following we reveal a simplification of the QSC in the BFKL regime, reminiscent of what happens for local operators, which allows us to go beyond a numerical analysis and derive analytic results up to wrapping corrections at $\sim g^{L+2}$.

\paragraph{QSC Generalities.}
Like the original Baxter equation \eqref{eq:LipatovBaxter} the QSC is a set of equations on Q-functions. However, unlike the LO BFKL, which has $2$ solutions, the exact QSC equations contains large number of Q-functions, which are no longer meromorphic functions of $u$, but the multi-valued functions with quadratic branch-cuts~(see \cite{Gromov:2017blm} for reviews). 

For context, the basic Q-functions are called $\bP_a(u)$ and $\bQ_i(u),\,a,i=1,\dots,4$. They are functions with a cut between $\pm 2g$. While $\bP_a$ are otherwise analytic, $\bQ_i$ has an additional infinite towers of cuts at $(-2g-\ii n,2g-\ii n),n\in \mathbb{Z}_{>0}$. Analytic continuation around the branch-points on the real axis, denoted by $\tilde{\bP}_a(u),\tilde{\bQ}_i(u)$, is a symmetry of the QSC acting as $\tilde {\bf Q}_i = \omega_{ij}\chi^{jk}{\bf Q}_k$ and $\tilde{\bP}_a = \mu_{ab}\chi^{bc} \bP_{c}$. The matrices $\omega_{ij},\mu_{ab}$ are both anti-symmetric and $\chi^{ij}=(-1)^{i}\delta^{i+j-5,0}$. 

\paragraph{Asymptotic Bethe Ansatz (ABA).}
For local operators at weak coupling the QSC simplifies significantly and the standard ABA equations emerges. Intriguingly, one can trace the scaling $\omega^{12}\sim g^{-2L-4} $ as the underlying reason for this simplification. In this limit the zeros $\mu_{12}(u_k+\frac{\ii}{2}) = 0$ are identified as momentum-carrying roots, corresponding to magnons with momentum $e^{\ii p_k}=x^+_k/x^-_k$ and energy $E_k=2\ii g/x^+_k -2\ii g/x^-_k$ with $g\;(x+1/x)=u$ the Zhukovsky variable or alternatively $E_k = \sqrt{1+16\,g^2 \sin^2{\frac{p_k}{2}}}$. 

\section{Asymptotic Baxter-Bethe Ansatz}
In the BFKL regime we find that $\omega^{13}$ dominates instead of $\omega^{12}$. This change effectively swaps the role of $\Delta$ and $S$. The scaling $\omega^{13} \sim g^{-L-2}$ implies similar simplification as in the case of the ABA for local operators, but with several novel features which we now describe. The zeros of $\mu_{12}$ once again play a pivotal role, for the first HTs we find that there are always $4$ zeros, $z_i$, and they are all located on the unit circle in the Zhukovsky plane, $\abs{z_i}=1$, and satisfy $z_3=-\frac{1}{z_2},z_4=-\frac{1}{z_1}$. Due to their novel location, the momenta of the corresponding modes now read $p_k =\frac{1}{\ii} \log z_k^2$ and the dispersion relation becomes
\begin{equation}\label{eq:OmegaDispersion}
    \omega =  \ii g \sum_{k=1}^{4}\left(z_k-\tfrac{1}{z_k}\right) =  2 g \sum_{k=1}^{2}\sin \frac{p_k}{2}\,.
\end{equation}

In this limit there is a particularly simple subset of Q-functions whose roots enter into the Bethe-like equations. A convenient choice is given by $\bP_1,\fQ_{1|1},\fQ_{1|13}$ and $\fQ_{12|13}$. We managed to
derive the asymptotic approximation of all these Q-functions, in particular $\bP_1 = \sigma_{\circ} x^{-\frac{L}{2}-1}$ where $\sigma_{\circ}$ is a solution of the ``crossing" equation 
\begin{equation}\label{eq:CrossingEq}
    \prod_{\pm}{\sigma}_{\circ}(u\pm i0) = \prod_{n=0}^{\infty} \left(\frac{\kappa(x)}{x^2}\right)^{[2+2n]}
    \left(\frac{\bar\kappa(x)}{x^2}\right)^{[-2-2n]}\;,
\end{equation}
where $\kappa(x)\equiv \prod_i(x-z_i)$ and 
$\bar\kappa(x)\equiv \prod_i(x-1/z_i)$. The solution to
\eq{eq:CrossingEq} can be written as a double integral,  similar
to the BES dressing phase~\eq{crossingsolve}.

The most difficult Q-function to fix is $\fQ_{1|1}$ which have an infinite number of zeros. This renders the standard Bethe equation technique futile, instead we fix $\fQ_{1|1}$ using a Baxter equation which generalizes \eqref{eq:LipatovBaxter}. The need to include a functional equation is the reason we refer to our approach as an Asymptotic Baxter-Bethe Ansatz (ABBA) as opposed to simply ABA. 

\paragraph{The Auxiliary Baxter Equation.}
We found that we can factorize $\fQ_{1|\beta}$ for $\beta=1,3$ into a fixed function with branch-cuts and a meromorphic function $\baxterQ_{1|\beta}$: $\fQ_{1|\beta} = \baxterQ_{1|\beta} \prod_{n=0}^{\infty}\frac{\kappa^{[2n+1]}_1}{\bar{\kappa}^{[2n+1]}_2}$ with $\kappa_i = (x-z_i)(x+\frac{1}{z_i})$. We also introduce the following rapidity parameterization $\theta_i =g\;(z_i+\frac{1}{z_i})$ to get
\begin{align}\label{eq:Baxter}
    ((u+&\tfrac{\ii}{2})^2-\theta_2^2) \, \mathbb{Q}_{1|\beta}(u+\ii) + ((u-\tfrac{\ii}{2})^2-\theta_1^2) \mathbb{Q}_{1|\beta}(u-\ii)\nonumber\\
    &=  \left(2u^2-\tfrac{1}{4}(\Delta^2+4 \theta_1^2+4\theta_2^2+1)\right)\, \mathbb{Q}_{1|\beta}(u)\,.
\end{align}
Remarkably, this equation can be solved exactly (see \eq{Q11sol}). For more general state we do not expect such a luxury. 

\paragraph{Middle-node equation for massless modes.}
In order to fix $z_i$'s themselves we have to use the gluing condition $\bQ_1(u+i0)=\alpha \bar\bQ_3(u-i0)$, which is a part of the QSC formalism. After a technically involved derivation, scratched in supplemented materials and to be presented in full in \cite{UpcommingLong}, the gluing condition reduces to 
\begin{equation}\label{eq:middlenode}
    1 \!= \!\frac{(\ii z_k)^{-2L-4}}{G^{\frac{1+\Delta}{2}}_{kk}G^{\frac{1-\Delta}{2}}_{kk}} \frac{\sigma_\circ^{2}(-z_k)}{\sigma_\circ^{2}(-\frac{1}{z_k})}\! \prod_{n\neq 0} \!\frac{\kappa(z_k^{[2n]})}{\bar{\kappa}(z_k^{[2n]})}    \!\prod_{l=1,2}\!\!\frac{G_{kl}^{\frac{1+\Delta}{2}}\!G_{kl}^{\frac{1-\Delta}{2}}\!}{G^{1}_{kl}}
\end{equation}
where $G_{kl}^{\delta} = \frac{\Gamma(\delta+\ii \theta_k+\ii \theta_l)\Gamma(\delta+\ii \theta_k-\ii \theta_l)}{\Gamma(\delta-\ii \theta_k-\ii \theta_l)\Gamma(\delta-\ii \theta_k+\ii \theta_l)}$. The above equation should also be supplemented with a selection rule $(z_1 z_2)^{L-2}\to 1$ as $g\to 0$. 
One can show that \eq{eq:middlenode} can be written in the form $(z_k)^{2L+4}=B(z_k)\prod_{l=1}^4 S(z_k,z_l)$, where $S$ satisfies $S(z_k,z_l)=1/S(z_l,z_k)$ (see \eq{eq:skl}).

\paragraph{Counting states.} When $g\to 0$ we have $\sigma_\circ\to 1$ and $\theta_k\to 0$ and thus \eqref{eq:middlenode}
simplifies to $1=(iz_k)^{-2L-4}$ or $z_k=-i e^{\frac{\pi i n_k}{L+2}},\;n_1=1,\dots,L+1$ with the selection rule imposing $n_1+n_2$ being even. Note that the other $2$ $z$'s are fixed by $z_4=-1/z_1$ and $z_3=-1/z_2$ (due to the parity of the states we consider). This gives $\left\lfloor \frac{L^2}{4}\right\rfloor$ for the  counting of the states as demonstrated in TABLE ~\ref{tab:my_label}. 
\begin{table}[h]
    \centering
    \begin{tabular}{c|l}
       L  & states \\ \hline
        2 & \adjustbox{valign=c}{\includegraphics[scale=0.3]{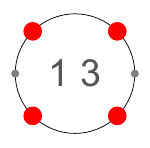}}\\
       3  & \adjustbox{valign=c}{\includegraphics[scale=0.3]{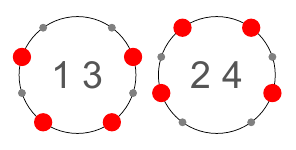}}\\
       4  & \adjustbox{valign=c}{\includegraphics[scale=0.3]{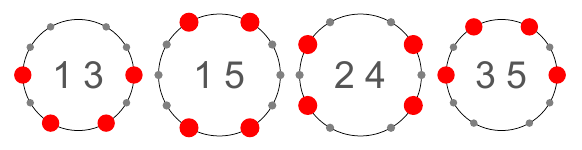}}\\
       5  & \adjustbox{valign=c}{\includegraphics[scale=0.3]{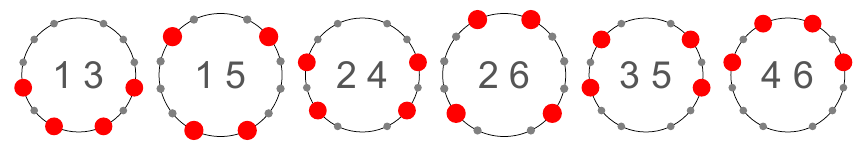}}\\
    \end{tabular}
    \caption{States for various R-charge $L$. Numbers inside the circle are $n_1,\;n_2$, red dots are $z_k$'s. Horizontal reflection is equivalent to $g\to -g$.}
    \label{tab:my_label}
\end{table}

\paragraph{Explicit results.}

We assume the ansatz $z_k=z_k^0+\sum g^n a_{k,n}$, with $i z_k^0$ a fixed root of unity, and plug it into the quantization condition \eqref{eq:middlenode}. After that we can compute $\omega(\Delta)$ from \eqref{pomeroneigenvalue}.
For example, for $L=3$ there are only two states, related by $g\to-g$. For the $(1,3)$ solution we find
\begin{equation}\label{eq:L3omega}
\omega\!=\!2 g-4  \chi g^2\!-\frac{2 \pi ^2 g^3}{3}+24\!\left(\!\frac{\pi^2}{18} \chi\! + \chi ''\!+\!\frac{7\zeta_3}{6} \!\!\right)\!g^4
\!+\mathcal{O}(g^{\!5})\,.
\end{equation}
Note that all terms has uniform transcedentality given by the order in $g$ minus $1$. We checked that this is true in general to order $g^{8}$ for general $L$.
Note that at order $g^{L+2}$ the ABBA approximation breaks down and one has to take into account ``wrapping"-like corrections. We see that \eqref{eq:L3omega} generalizes the Pomeron eigenvalue \eqref{pomeroneigenvalue}. In particular for $\Delta=0$ we find
$
\omega(0)\simeq 2 g+8 \log 4 g^2-\frac{2 \pi ^2 g^3}{3}-\frac{4}{3} \left(\pi ^2 \log 16+105 \zeta_3\right) g^4
$, which agrees perfectly with the leading term analytically and numerical results for other terms of~\cite{Klabbers:2023zdz}.
For larger $L$ the predictive power of our approach increases since \eqref{eq:OmegaDispersion} is valid up to and including $g^{L+1}$. We present results for $L=4$ in~\eq{L4explicit}. The result for any $L$ up to $g^2$ is given in \eqref{demoeq} and up to $g^4$ in \eq{moreterms}. Further explicit results can be found in the attached {\it Mathematica} notebook.

\section{Branching Regime}
\begin{figure}[t]
    \centering
    \includegraphics[width=\columnwidth]{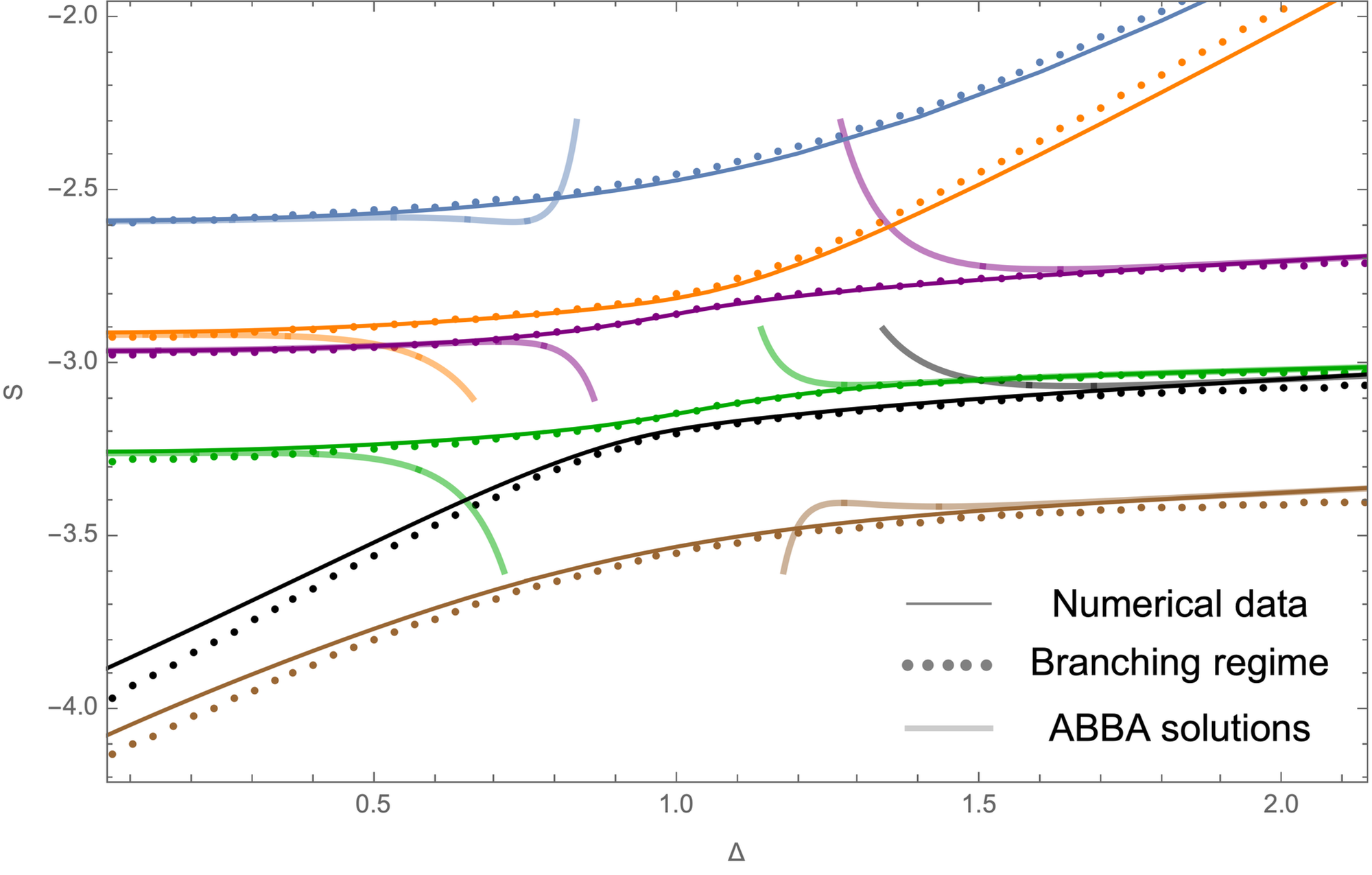}
    \caption{Numerical data at $g=1/10$ for $L=4$ (solid line) compared with the ABBA result at $g^5$ order~\eq{L4explicit} (pale line) and the leading order {\it branching} equation~\eq{curveL4} (dotted line). See also FIG.~\ref{fig:branchhigherL} for higher $L$'s.}
    \label{fig:branching}
\end{figure}
The ABBA allows us to obtain the HTs to a high perturbative order. In this section we study how they merge with the DTs. Let us start with $L=3$, we know from numerics that there are $2$ HTs mixing with one diagonal one. In order to study the mixing between these $3$ branches it is natural to zoom to the corner by introducing the rescaled variables $\omega = \Omega\; g$ and $\Delta=1+g \;{\mathbb D}$. In this scaling our result~\eq{eq:L3omega} for the the two HTs become
\begin{equation}\label{Omega12}
\Omega_{1,2}\simeq \pm 2-\frac{8}{\mathbb D}+\frac{96}{\mathbb D^3}+{\cal O}\left(\frac{1}{{\mathbb D}^4}\right)+{\cal O}(g^2)\;.  
\end{equation}
As there are $3$ trajectories $\omega(\Delta)$ mixing at this point the full Riemann surface reduces to just a degree $3$ polynomial $\Omega^3+A\Omega^2+B\Omega+C=0$ with coefficients $A,B,\;C$ non-singular functions of ${\mathbb D}$. Furthermore at large ${\mathbb D}$ we should require that two solutions satisfy \eqref{Omega12}, and the third one is $\Omega_3\simeq {\mathbb D}$, fixing uniquely
\begin{equation}\la{curveL3}
\Omega ^3-{\mathbb D} \Omega ^2-20 \Omega +4 {\mathbb D}=0\;,
\end{equation}
which reproduces the branching regime of FIG.~3 in \cite{Klabbers:2023zdz}.
The formula \eqref{curveL3}, albeit very simple, is quite powerful. Solving for $\mathbb{D}$ and restoring $\omega$ we can find $\gamma(g)=\Delta-1-\omega$:
\begin{equation}
-\frac{16 \omega  g^2}{\omega ^2-4 g^2}\simeq
-\frac{16 g^2}{\omega }-\frac{64 g^4}{\omega ^3}-\frac{256 g^6}{\omega ^5}-\frac{1024 g^8}{\omega ^7}
\end{equation}
which reproduces correctly the divergence of $\Delta$ and $\omega=0$ for the known $4$-loop result \cite{Kotikov:2007cy,Beccaria:2007cn}, as well as gives prediction to all higher loops in agreement with the guess in \cite{Kotikov:2007cy}. At the same time, one can also give prediction for all terms in the $1/{\mathbb D}$ expansion in \eqref{Omega12} as a solution of cubic equation, meaning that it resums all $\frac{g^n}{(\Delta-1)^k}$ singularities on HTs. Furthermore, one can try to repeat this exercise at $\Delta\to 2k-1$ for $k=1,2,\dots$. We found that that amounts to the simple change $\Delta=(2k-1)+g{\mathbb D}$ but otherwise gives the same curve~\eq{curveL3}. We expect that corrections in $g$ will be different.

Applying the same logic, and assuming that there are $\lfloor \frac{L}{2}\rfloor$ diagonal trajectories mixing with $\lfloor \frac{L^2}{4}\rfloor$ the horizontal curves as indicated by numerics for $L\leq 4$, it is possible to build curves for any $L$. For example, for $L=4$ we get the following curve
\begin{align}\la{curveL4}
    \Omega ^6&-2 \mathbb{D} \Omega ^5+\left(\mathbb{D}^2-36\right) \Omega ^4+48 \mathbb{D} \Omega ^3\\
\nonumber    &+\left(192-12 \mathbb{D}^2\right) \Omega ^2-128 \mathbb{D} \Omega -256=0\;,
\end{align}
which gives a surprisingly good approximation even far from the branching regime, see FIG.~\ref{fig:branching}. The curve
\eq{curveL4} also gives a prediction for the leading singularity for the two DTs
\begin{equation}\label{branchingsp4}
\Delta^{L=4}_{\pm}-1\simeq\omega -4g^2\frac{ 3\pm\sqrt{5}}{\omega }-16g^4\frac{25\pm11 \sqrt{5}}{5 \omega ^3}+\dots.   
\end{equation}
It would be interesting to check this prediction by solving the $\mathfrak{sl}(2,{\mathbb R})$ Hamiltonian for light-rays \cite{Caron-Huot:2016tzz}. 
Let us make a curious observations: Firstly, the number of trajectories which branch from diagonal to horizontal at $\omega\simeq 0$ is equal to the number of local operators with spin $2$. In fact, from the available numerical data we find that the DTs that mix are exactly those that contain local operators with spin $2$, so we conjecture that this is true in general.
Secondly, the residues $b_n$ at the leading poles $b_n g^2/\omega$,
which we call {\it branching spectrum}, are related in a simple way to the 1-loop anomalous dimension of these operators with $S=2$. For example, we recognize the coefficient of $\frac{g^2}{\omega}$ in \eq{branchingsp4} as almost the anomalous dimension of $\tr D^2 Z^4$! We checked that in general one has $b_n=32-2\gamma_n^{(1)}$ for all $3\leq L\leq 11$ (see \eq{onlocal}). This rather mysterious relation between residues and the anomalous dimensions of local operators deserves further study. In particular, one can ask if there is a similar relation for $S=4$ and poles at $S\simeq -L-1$ etc. Finally, there is a lot of data to be explored in the future in terms of $g^2$ corrections to the branching curves, which can be deduced from the ABBA.

\section{Conclusions}
In this letter we presented the Asymptotic Baxter-Bethe Ansatz for the first HTs for general parity symmetric $\mathfrak{sl}(2)$ operators.
Whereas the equations can be rigorously derived starting from QSC, it would be interesting to understand the linear and quadratic terms 
in $g$ from a field theory perspective as well as uncovering the underlying spin-chain. The framework of BK-JIMWLK~\cite{Jalilian-Marian:1996mkd,Jalilian-Marian:1997jhx,Iancu:2001ad}, evolution seems to be the most natural to address these questions.
The higher order terms in $g$ will hopefully prove helpful to fix higher order terms in the linearized BFKL evolution Hamiltonian in ${\cal N}=4$ SYM
and in QCD via the Kotikov-Lipatov maximal transcedentality principle.

Our results can be further generalized to other sectors and to non-parity symmetric operators as well as subleading horizontal trajectories \cite{UpcommingLong}.

In order to form a complete picture of the Riemann surface of Regge trajectories it would be interesting to study the corner points where 
the BFKL and DGLAP trajectories meet and where the QSC may simplify as well.

Whereas we have clear evidence that effective massless degrees of freedom emerge in this limit, the physics of these modes is not yet clear. Perhaps intuition can be gained from the AdS${}_3$/CFT${}_2$ QSC \cite{Ekhammar:2021pys,Cavaglia:2021eqr} where our techniques looks relevant to describe similar massless modes.

Finally, the regime $L\to\infty$ may have some interesting quasi-classical description, unveiling which can give a new perspective on the BFKL regime.

\begin{acknowledgments}
\section{acknowledgments}
 We are grateful to F.~Levkovich-Maslyuk and G.~Sizov for collaboration at the initial stage of the project as well as to M.~Alfimov, I.~Balitsky, B.~Basso, S.~Caron-Huot, V.~Kazakov, R.~Klabbers, P.~Kravchuk, G.~Korchemsky, J.~Mann, L.~Preti, A.~Sabio Vera, I.M.~Sz\'{e}cs\'{e}nyi, E.~Sobko and P.~Vieira
 for fruitful discussions. 
N.G. is grateful to LPENS Paris for warm hospitality while a part of this work was done. 
N.G. is grateful to M.~Nefedov for organising the motivational workshop ``Workshop on overlap between QCD resummations" and the organizers of ``Bootstrap Meets Integrability" for creating a productive environment.
The work of N.G. and S.E. was
supported by the European Research Council (ERC) under the European Union’s Horizon 2020 research and innovation program – 60 – (grant agreement No. 865075) EXACTC.  The work of M.P. is supported by Marie Skłodowska-Curie Global Fellowship
(HORIZON-MSCA-2022-PF-01) BOOTSTRABILITY-101109934 and by the INFN SFT specific initiative.
\end{acknowledgments}

\bibliography{ref}
\onecolumngrid
\newpage
\setcounter{equation}{0} 
\renewcommand{\theequation}{S\arabic{equation}} 
\setcounter{secnumdepth}{2}
\section{The Dressing Phase and Effective ABA}
An important part in the ABA is the Beisert-Eden-Staudacher dressing phase \cite{Beisert:2006ez}, an overall factor in the S-matrix that cannot be fixed by the Yang-Baxter equation. Instead it must be found through a crossing equation first derived in the context of AdS/CFT integrability by Janik \cite{Janik:2006dc}. To be precise, in the $\algsl(2)$ sector the ABA becomes
\begin{equation}\label{eq:sl2BAE}
    \left(\frac{x^+_i}{x^-_i}\right)^{L} = \prod_{\substack{ j\neq i}} S(u_i,u_j)\;,
    \quad
    S(u,v) = \frac{x^-_u - x^+_v}{x^+_u - x^-_v} \frac{1-\frac{1}{x^+_u x^-_v}}{1-\frac{1}{x^-_u x^+_v}} \sigma^2_{\text{BES}}(u,v)\;,
\end{equation}
where $\sigma_{\text{BES}}(u,v) = \frac{\sigma_{\bullet}(u^+,v)}{\sigma_{\bullet}(u^-,v)}$ is the dressing phase and the factor $\sigma_{\bullet}$ is given as 
\begin{equation}\label{eq:SigmaMassive}
    \log \sigma_{\bullet}(u,v) = \oint \frac{dz}{2\pi \ii} \oint \frac{dy}{2\pi \ii} \frac{1}{x(u)-z}\log\left( \frac{\Gamma\left(1+\ii u_{z}-\ii u_{y}\right)}{\Gamma\left(1-\ii u_{y}+\ii u_{z}\right)}\right)\left(\frac{1}{x(v-\tfrac{i}{2})-y}-\frac{1}{x(v+\tfrac{i}{2})-y}\right) \,.
\end{equation}

In this section we discuss how to express our dressing phase building block $\sigma_{\circ}$ and Bethe equation \eqref{eq:middlenode} in forms reminiscent of \eqref{eq:sl2BAE} and \eqref{eq:SigmaMassive}.

\subsection{Solving Crossing Equation}
In the QSC formalism the crossing equation emerges in the asymptotic or weak coupling regime from the $\bP\mu$-system \cite{Gromov:2014caa}. In our setting the same happens, but the crossing equation is now different due to the novel massless momentum carrying roots. The crossing equation \eq{eq:CrossingEq}, repeated here for convenience, reads
\begin{equation}\label{eq:CrossingSM}
    \sigma_{\circ}(z+\ii 0)\, \sigma_{\circ}(z-\ii 0) = \prod_{n=0}^{\infty} \left(\frac{\kappa}{x^2}\right)^{[2+2n]}\left(\frac{\bar{\kappa}}{x^2}\right)^{[-2-2n]}\,.
\end{equation}
From the $\bP\mu$ system and the asymptotics of Q-functions it follows that $\bP_{1} \propto x^{-\frac{L}{2}-1}\sigma_{\circ}$. The QSC-requires that $\bP_{1} \sim_{u\rightarrow \infty}  u^{-\frac{L}{2}-1}$ and that $\bP_1$ is an analytic function outside of a cut $(-2g,2g)$. This implies that $\sigma_{\circ}$ is analytic on the main sheet and $\sigma_{\circ}(\infty) = 1$. With this input the solution to \eqref{eq:CrossingSM} is uniquely fixed to be given by the following double integral
\beqa\label{crossingsolve}
\log\sigma_{\circ}(z)=\!\sum_{i=1}^4\!\left[\frac{1}{2}\sum_{n\neq 0} \log\left(\!{1-\frac{1}{z\;x(\theta_i+in)}}\right)\!+\frac{1}{2}\!\oint \!\frac{dx}{2\pi i}\oint\!
\frac{dy}{2\pi i}
\frac{1}{z-x}
\left(\frac{1}{y - z_i} - \frac{1}{y - 1/z_i}\right)
\log \frac{\Gamma (1+i u_x-i u_y)}{\Gamma (1-i u_x+i u_y)}\right]
\eeqa
where the integrals over a circle of the radius $1-\epsilon$ for some small $\epsilon$, needed for regularization. We see that this is indeed similar to \eqref{eq:SigmaMassive}. 

\subsection{Effective Bethe Ansatz Equation}
In this section we rewrite the effective Bethe-ansatz equation \eq{eq:middlenode} using \eq{crossingsolve} to reach a form closer to \eqref{eq:sl2BAE}. Let us duplicate the Bethe equation for $z_i$ as presented in the main text here:
\begin{equation}\label{eq:middlenode2}
    1 \!= \!\frac{(\ii z_k)^{-2L-4}}{{\bf G}^{\frac{1+\Delta}{2}}(\theta_k,-\theta_k){\bf G}^{\frac{1-\Delta}{2}}(\theta_k,-\theta_k)} \frac{\sigma_{\circ}^{2}(-z_k)}{\sigma_{\circ}^{2}(-\frac{1}{z_k})}\! \prod_{n\neq 0} \frac{\kappa(z_k^{[2n]})}{\bar{\kappa}(z_k^{[2n]})}    \!\prod_{l=1}^4\!\!\frac{{\bf G}^{\frac{1+\Delta}{2}}(\theta_k,\theta_l){\bf G}^{\frac{1-\Delta}{2}}(\theta_k,\theta_l)}{{\bf G}^{1}(\theta_k,\theta_l)}
\end{equation}
where
${\bf G}^\delta(\theta_k,\theta_l) = \frac{\Gamma(\delta+\ii \theta_k-\ii \theta_l)}{\Gamma(\delta-\ii \theta_k+\ii \theta_l)}$. We emphasize that for more general states we expect a nested system of Bethe-Baxter equations to describe the spectrum. However, for the current states we were lucky to be able to solve the Baxter equation analytically, which then allows us to write the effective Bethe Ansatz involving only the massless momentum carrying nodes $z_i$.

Next, we notice that the combination of $\sigma_{\circ}$'s in \eq{eq:middlenode2} can be written in a more standard form
\beqa
\left(\frac{\sigma_{\circ}(-z_k)}{\sigma_{\circ}(-1/z_k)}\right)^2
\prod_{n\neq 0} \frac{\kappa(z_k^{[2n]})}{\bar{\kappa}(z_k^{[2n]})}
=\prod_{l=1}^4\left[\Sigma_{\circ}(z_k,z_l)
K(z_k,z_l)\right]
\eeqa
where we defined
\beqa
\log\Sigma_{\circ}(z_k,z_l)\equiv 
-\oint_{|x|<1} \frac{dx}{2\pi i}\oint_{|y|<1}
\frac{dy}{2\pi i}
\left(\frac{1}{1/z_k-x}-\frac{1}{z_k-x}\right)
\log \left(\frac{\Gamma (+i u_x-i u_y+1)}{\Gamma (-i u_x+i u_y+1)}\right)
\left(\frac{1}{1/z_l-y}-\frac{1}{z_l-y}\right)\;,
\eeqa
and
\beq
K(z_k,z_l)\equiv 
\prod_{n\neq 0}\frac{x(\theta_l+in)-\frac{1}{z_k}}{x(\theta_l+in)-{z_k}}
\frac{x(\theta_k+i n)-z_l}{x(\theta_k+i n)-\frac{1}{z_l}}\;.
\eeq
Finally we combine all pieces together and define a scalar ``S-matrix"
\beq\la{eq:skl}
S(z_k,z_l)=K(z_k,z_l)\Sigma_\circ(z_k,z_l) \frac{{\bf G}_{kl}^{\frac{1+\Delta}{2}}{\bf G}_{kl}^{\frac{1-\Delta}{2}}}{{\bf G}_{kl}^l}\;\;,\;\;S(z_k,z_l)=\frac{1}{\bar S(1/z_k,1/z_l)}=\frac{1}{S(z_l,z_k)}\;,
\eeq
in terms of which \eq{eq:middlenode} becomes 
\beq
(i z_{k})^{2L+4} =B(z_k)\prod_{l=1}^4 S(z_k,z_l)
\eeq
where $B(z_k)\equiv 1/\left[{\bf G}^{\frac{1+\Delta}{2}}(\theta_k,-\theta_k){\bf G}^{\frac{1-\Delta}{2}}(\theta_k,-\theta_k)\right]$. Clearly this form bears a striking resemblance to \eqref{eq:sl2BAE}, it would be interesting to understand if this equation could be derived starting from either a spin chain or from a bootstrap approach using only symmetry.
\subsection{Perturbative Expansion}
The dressing phase (or more precisely the building block of the dressing phase) \eqref{eq:middlenode2} can be easily computed by residues at each order in $g$. The result reads
\begin{align}
\nonumber\log\sigma_{\circ}(x)&=\sum_{i=1}^4 \left(
\left(\frac{z_i}{x}-\frac{1}{2 x^2}+\frac{1}{x z_i}\right) (ig)^2\zeta_2+\left(-\frac{z_i^2}{x}+\frac{z_i}{x^2}-\frac{1}{x^2 z_i}+\frac{1}{x z_i^2}\right) (ig)^3\zeta_3\right.\\
&+\left(\frac{z_i^3}{x}-\frac{3 z_i^2}{2 x^2}+\frac{\left(6 x^2+1\right) z_i}{x^3}-\frac{16 x^2+1}{4 x^4}+\frac{6 x^2+1}{x^3 z_i}-\frac{3}{2 x^2 z_i^2}+\frac{1}{x z_i^3}\right) (i g)^4\zeta_4\\
\nonumber&\left.+\left(-\frac{z_i^4}{x}+\frac{2 z_i^3}{x^2}-\frac{2 \left(5 x^2+1\right) z_i^2}{x^3}+\frac{\left(10 x^2+1\right) z_i}{x^4}+\frac{-10 x^2-1}{x^4 z_i}+\frac{2 \left(5 x^2+1\right)}{x^3 z_i^2}-\frac{2}{x^2 z_i^3}+\frac{1}{x z_i^4}\right) (ig)^5\zeta_5
\right)+O\left(g^6\right)
\end{align}
\section{Solution of the Baxter equation}\label{sp:SolutionBaxter}
In this section we present the solution of the Baxter equation \eq{eq:Baxter}. 
To solve this equation one performs a Mellin transform in $u$ variable, reducing it to a second order differential equation. This differential equation can be solved and transforming back we get the following solution, which we can directly verified as a solution to \eq{eq:Baxter},
\beqa\label{Gsol}
&&G^{1}(\Delta,\theta_1,\theta_2,u)\equiv\\
\nonumber&&\frac{(-1)^{-1/4} \;2^{1-\Delta }\; \frac{e^{\frac{3 i \pi  \Delta }{4}}}{
1+e^{i \pi  \Delta }
}\; \sqrt{\pi } \ \Gamma \left(1-\frac{\Delta }{2}\right) \Gamma \left(-i u+i \theta_1+\frac{1}{2}\right) \Gamma \left(-2 i \theta_2\right) }{ \Gamma \left(-i u-i \theta_2+\frac{1}{2}\right) \Gamma \left(-\frac{\Delta }{2}-i \theta_1-i \theta_2+\frac{1}{2}\right) \Gamma \left(-\frac{\Delta }{2}+i \theta_1-i \theta_2+\frac{1}{2}\right) \Gamma \left(i \theta_1+i \theta_2+1\right)}\\
\nonumber&&
\, _3F_2\left(i u+i \theta_2+\frac{1}{2},-\frac{\Delta }{2}+i \theta_1+i \theta_2+\frac{1}{2},\frac{\Delta }{2}+i \theta_1+i \theta_2+\frac{1}{2};i \theta_1+i \theta_2+1,2 i \theta_2+1;1\right)\;.
\eeqa
The Baxter equation \eq{eq:Baxter} has two linearly independent solutions, one associated with $\fQ_{1|1}$ and another with $\fQ_{1|3}$, similar to \cite{Alfimov:2014bwa}. In order to generate the second solution we notice that 
\eq{eq:Baxter} is invariant under $\theta_1\to -\theta_1$ and $\theta_2\to - \theta_2$.
However, \eq{Gsol} satisfies a non-trivial identity $G^{(1)}(\Delta,\theta_1,\theta_2)=G^{(1)}(\Delta,-\theta_1,\theta_2)$. So the second solution can be obtained by changing the sign of both $\theta$'s. In order to get the $\fQ_{1|1}$ we can make linear combinations of 
$G^{(1)}(\Delta,\theta_1,\theta_2)$ and 
$G^{(1)}(\Delta,-\theta_1,-\theta_2)$, which has pure power like asymptotics $u^{\frac{\Delta-1}{2}}$.
We find
\beqa
{\mathbb Q}_{1|1}=G^{(1)}(+\Delta,+\theta_1,+\theta_2,u)+G^{(1)}(+\Delta,-\theta_1,-\theta_2,u)\;.
\eeqa
In order to perform gluing procedure of QSC,
we also need to know ${\mathbb Q}_{1|3}$. It can also be obtained as a linear combination of the same blocks, but instead we can use the $\Delta\to-\Delta$ symmetry of the equation to immediately get
\beqa
{\mathbb Q}_{1|3}=G^{(1)}(-\Delta,+\theta_1,+\theta_2,u)+G^{(1)}(-\Delta,-\theta_1,-\theta_2,u)\;.
\eeqa
Obtained in this way solutions has poles in the lower half plane at $\pm\theta_1-i/2-i n$, $n=0,1,2,\dots$. We need to know the residue of the first poles at $n=0$.
This pole of 
$G^{(1)}(\Delta,\theta_1,\theta_2,u)$ at $u=+\theta_1-i/2$ can be computed explicitly to be 
\beq\nonumber
r^{(1)}(\Delta,\theta_1,\theta_2)=\frac{(-1)^{3/4} e^{-\frac{1}{4} i \pi  \Delta } \left(e^{i \pi  \Delta +2 \pi  \theta _1}+e^{2 \pi  \theta _2}\right) \left(e^{2 \pi  \left(\theta _1+\theta _2\right)}-1\right) \Gamma (1-\Delta ) \Gamma \left(\frac{\Delta +1}{2}\right)}{\left(e^{4 \pi  \theta _1}-1\right) \left(e^{4 \pi  \theta _2}-1\right) \Gamma \left(2 i \theta _1+1\right) \Gamma \left(\frac{1}{2}-i \theta _1-i \theta _2-\frac{\Delta }{2}\right) \Gamma \left(\frac{1}{2}-i \theta _1+i \theta _2-\frac{\Delta }{2}\right)}
\eeq
and the one at $u=-\theta_1-i/2$ is $r(\Delta,-\theta_1,\theta_2)$.

\section{Additional QSC Details}

\subsection{ABBA Q-functions}
When constructing the full ABBA we need to find all Q-functions along a Hasse path. A possible, and especially convenient choice for us, is to pick $\{\bP_1,\fQ_{1|1},\fQ_{1|13},\fQ_{12|13}\}$. We already fixed $\fQ_{1|1}$ using the Baxter equation \eqref{eq:Baxter}, the remaining Q-functions can be found following the procedure outlined in \cite{Gromov:2014caa}, this gives
\begin{equation}\label{eq:ABBAQfunctions}
    \bP_{1} \propto \sigma_{\circ} \, x^{-\frac{L}{2}-1}\;,
    \quad
    \fQ_{1|\beta} \propto \betheQ_{1|\beta} \, \prod_{n=0}^{\infty} \frac{\kappa^{[2n+1]}_1}{\bar{\kappa}^{[2n+1]}_2}\;,
    \quad
    \fQ_{1|13} \propto \frac{x^{\frac{L}{2}+3}}{\sigma_{\circ} \bar{\kappa}}\prod_{n=0} \left(\frac{\kappa^{[2+2n]}}{\bar{\kappa}^{[2+2n]}}\right)\;,
    \quad
    \fQ_{12|13} \propto \prod_{n=0}^{\infty} \frac{\kappa^{[2n+1]}}{\bar{\kappa}^{[2n+1]}}\;.
\end{equation}
We recall $\kappa = \prod_{i=1}^{4}(x-z_i)$, $\bar{\kappa}=\prod_{i=1}^{4}(x-\frac{1}{z_i})$ and $\kappa_{i} = (x-z_i)(x+\frac{1}{z_i})$. 

\subsection{Derivation of the middle node equation}
The gluing equations relates the analytic continuation of $\bQ_{i}$ with $\bar{\bQ}_i$ through the so-called \emph{gluing conditions}. For our state the relevant gluing equation becomes
\beq\label{eq:GluingQ13}
    \tilde{\bQ}_1 = \alpha\, \bar{\bQ}_{3}\;,
    \quad
    \tilde{\bQ}_{3} = \frac{1}{\bar{\alpha}} \bar{\bQ}_{1}\;.
\eeq
Let us for simplicity introduce a subscript $\beta=1,3$. QQ-relations dictate that $\bQ_{\beta} = \frac{\fQ_{1|\beta}^+-\fQ_{1|\beta}^-}{\bP_{1}}$ and using \eqref{eq:ABBAQfunctions} this gives $\bQ_{\beta}$ as
\begin{equation}\label{eq:bQFrombPQai}
    \bQ_{\beta} \propto
 \left(\prod_{n=0}^{\infty}\frac{\kappa^{[2n+2]}_{1}}{\bar{\kappa}^{[2n+2]}_{2}}\right)
 (gx)^{\frac{L+2}{2}}
 \frac{\mathbb{Q}_{1|\beta}^+ - \frac{\kappa_{1}}{\bar{\kappa}_{2}} \mathbb{Q}_{1|\beta}^- }{\sigma_{\circ}} \,,
 \quad
    \bar{\bQ}_{\beta} \propto \left(\prod_{n=0}^{\infty}\frac{\bar{\kappa}^{[-2n-2]}_{1}}{\kappa^{[-2n-2]}_{2}}\right)
    (g x)^{\frac{L+2}{2}}
    \frac{\bar{\mathbb{Q}}^-_{1|\beta} - \frac{\bar{\kappa}_{1}}{\kappa_{2}} \bar{\mathbb{Q}}_{1|\beta}^+ }{\sigma_{\circ}}\,.
\end{equation}
Note that $
(g x)^{\frac{L+2}{2}}\simeq u^{\frac{L+2}{2}}+\dots+g^{\frac{L+2}{2}}+\frac{g^{L+2}}{u^{\frac{L+2}{2}}}+\text{more singular}$ so that this term is regular up to the wrapping order $g^{L+2}$. We also have for the inverted argument
\begin{equation}
\begin{split}
    \bar{\bQ}_{\beta}\left(\frac{1}{x}\right) & \propto \left(\prod_{n=0}^{\infty}\frac{\bar{\kappa}^{[-2n-2]}_{1}}{\kappa^{[-2n-2]}_{2}}\right)
    g^{{L+2}}
    \frac{\bar{\mathbb{Q}}^-_{1|\beta} - \frac{{\kappa}_{1}}{\bar\kappa_{2}} \bar{\mathbb{Q}}_{1|\beta}^+ }{(gx)^{+\frac{L+2}{2}} \tilde{\sigma}_{\circ}}\,.
\end{split}
\end{equation}
As we can see in \eq{eq:bQFrombPQai} the r.h.s. contains poles at the 
branch cut, which are not allowed in the exact Q-functions. This is due to the fact that
the equation \eq{eq:bQFrombPQai} is valid on the top sheet at $u\sim 1$ with the $g^{L+1}$ precision only, so the poles we see is the artifacts of the approximation.
This seems to be a serious problem for the gluing as the gluing equation requires $\tilde {\bf Q}_1$ and in order to analytically continue we have to cross through the cut - the area where the approximation is no longer valid.
However, we can interpret the gluing condition \eqref{eq:GluingQ13} as a relation between the Fourier modes, i.e. expand $\bQ$ as $\sum c_n x^n$ around the cut, then \eqref{eq:GluingQ13} tells us that we have to match $c_{-n}^{{\bf Q}_1}$ with 
$\alpha \bar c_{n}^{{\bf Q}_3}$, that is $\oint_{\epsilon} \frac{dx}{x} \,x^{n}\,\bQ_1(x)-\alpha\,\oint_{-\epsilon} \frac{dx}{x} \,x^{n}\, \bar{\bQ}_3(\frac{1}{x}) = 0$ where $\epsilon$ denotes a infinitesimal shift of the radius of the contour. The shift in the contour allows us to deform it without crossing the cut to the domain where the asymptotic expressions \eq{eq:GluingQ13} are valid and use them to replace the exact $\bQ$'s under the integral.
We can now collapse the integrals and neglect all terms except those coming from singularities on the unit circle, as the contribution of the integral on the next sheet is suppressed by $g^{L+2}$. Since we have many more choices of $n$ as compared to number of singularities the only consistent way to make sure that all relations are satisfied is to set residues equal (with opposite sign). More precisely we find
\beq
\text{Res}_{x=-z_1}\left(
\alpha
\left(\prod_{n=0}^{\infty}\frac{\bar{\kappa}^{[-2n-2]}_{1}}{\kappa^{[-2n-2]}_{2}}\right)
    g^{L+2}
    \frac{\frac{{\kappa}_{1}}{\bar\kappa_{2}} \bar{\mathbb{Q}}_{1|3}^+ }{(gx)^{+\frac{L+2}{2}} \tilde{\sigma}_{\circ}}
+\left(\prod_{n=0}^{\infty}\frac{\kappa^{[2n+2]}_{1}}{\bar{\kappa}^{[2n+2]}_{2}}\right)
 (gx)^{\frac{L+2}{2}}
  \frac{\frac{\kappa_{1}}{\bar{\kappa}_{2}} \mathbb{Q}_{1|1}^- }{\sigma_{\circ}}\right) = 0\,.
\eeq
The residues can be computed from the explicit expression given in Section~\ref{sp:SolutionBaxter} to be
\beqa
&&\frac{\res{
\overline{\frac{\kappa_{1}(1/x)}{\bar{\kappa}_{2}(1/x)} \mathbb{Q}_{1|3}^-}
}{x=-z_1}}{\res{
\frac{\kappa_{1}}{\bar{\kappa}_{2}} \mathbb{Q}_{1|1}^-
}{x=-z_1}}=
\frac{4^{\Delta } \Gamma \left(\frac{\Delta }{2}+1\right) \Gamma \left(1-2 i \theta _1\right) \Gamma \left(-\frac{\Delta }{2}+i \theta _1-i \theta _2+\frac{1}{2}\right) \Gamma \left(-\frac{\Delta }{2}+i \theta _1+i \theta _2+\frac{1}{2}\right)}{\Gamma \left(1-\frac{\Delta }{2}\right) \Gamma \left(2 i \theta _1+1\right) \Gamma \left(\frac{\Delta }{2}-i \theta _1-i \theta _2+\frac{1}{2}\right) \Gamma \left(\frac{\Delta }{2}-i \theta _1+i \theta _2+\frac{1}{2}\right)}\;,\\
&&\frac{\res{
\overline{\frac{\kappa_{1}(1/x)}{\bar{\kappa}_{2}(1/x)} \mathbb{Q}_{1|3}^-}
}{x=1/z_1}}{\res{
\frac{\kappa_{1}}{\bar{\kappa}_{2}} (\mathbb{Q}_{1|1})^-
}{x=1/z_1}}=
\frac{4^{\Delta } \Gamma \left(\frac{\Delta }{2}+1\right) \Gamma \left(2 i \theta _1+1\right) \Gamma \left(-\frac{\Delta }{2}-i \theta _1-i \theta _2+\frac{1}{2}\right) \Gamma \left(-\frac{\Delta }{2}-i \theta _1+i \theta _2+\frac{1}{2}\right)}{\Gamma \left(1-\frac{\Delta }{2}\right) \Gamma \left(1-2 i \theta _1\right) \Gamma \left(\frac{\Delta }{2}+i \theta _1-i \theta _2+\frac{1}{2}\right) \Gamma \left(\frac{\Delta }{2}+i \theta _1+i \theta _2+\frac{1}{2}\right)}\;.
\eeqa
This gives the equation
\beq\label{eq:BAEV1}
-\alpha
 x^{-L-2}
\frac{\sigma_{\circ}}{\tilde{\sigma}_{\circ}}
{\left(\prod_{n=0}^{\infty}\frac{\bar{\kappa}^{[-2n-2]}_{1}}{\kappa^{[-2n-2]}_{2}}
\frac{\bar{\kappa}^{[2n+2]}_{2}}{\kappa^{[2n+2]}_{1}}
\right)
    }R(x) =1
\eeq
with $x=\{-z_1,\frac{1}{z_1}\}$ and $R(y) =\frac{\res{
\overline{\frac{\kappa_{1}(1/x)}{\bar{\kappa}_{2}(1/x)} (\mathbb{Q}_{1|3}^{(1)})^-}
}{x=y}}{\res{
\frac{\kappa_{1}}{\bar{\kappa}_{2}} (\mathbb{Q}_{1|1}^{(1)})^-
}{xy}}$. 

To get rid of $\alpha$ we simply divide \eqref{eq:BAEV1} for $x=-z_1$ with the same equation at $x=\frac{1}{z_1}$. This gives us, after some algebraic manipulations,
\beqa
1&=&(i z_k)^{-2L-4}
\frac{\sigma^2(-z_k)}{{\sigma^2(-\tfrac{1}{z_k})}}
\prod_{\pm}
\frac
{
\Gamma(\tfrac{1\pm\Delta}{2}-2i\theta_k)
}
{
\Gamma(\tfrac{1\pm\Delta}{2}+2i\theta_k)
}
\\
\nonumber&\times&\prod_{n\neq 0}\prod_{j=1,2}\frac{(z^{[-2n]}_k-z_j)(z^{[-2n]}_k+\tfrac1{z_j})}{
(z^{[+2n]}_k-\tfrac{1}{z_j})(z^{[+2n]}_k+z_j)
}\\
\nonumber&\times&
\prod_{j=1,2}\frac{\Gamma(-i\theta_k+i\theta_j+1)\Gamma(-i\theta_k-i\theta_j+1)}{
\Gamma(+i\theta_k-i\theta_j+1)\Gamma(+i\theta_k+i\theta_j+1)
}
\prod_{\pm}
\frac{
\Gamma(\tfrac{1\pm\Delta}{2}+i\theta_k-i\theta_j)
\Gamma(\tfrac{1\pm\Delta}{2}+i\theta_k+i\theta_j)
}{
\Gamma(\tfrac{1\pm\Delta}{2}-i\theta_k+i\theta_j)
\Gamma(\tfrac{1\pm\Delta}{2}-i\theta_k-i\theta_j)
}\;.
\eeqa
These are the Bethe equations \eqref{eq:middlenode} reported in the main text. By dividing the equations we lost some information - namely the relative sign, which can be restored by sending $g\to 0$, resulting in the selection rule $(z_1 z_2)^{L+2}\to 1$ at $g\to 0$ (implying that $n_1+n_2$ is even, which we used in the main text). 
We have in this section only derived the equation for $z_1$, to find the equation for $z_2$ one performs an analogous calculation but starts instead from the decomposition $\fQ_{1|\beta} = \hat{\betheQ}_{1|\beta} \,\prod_{n=0}^{\infty} \frac{\kappa^{[2n+1]}_2}{\bar{\kappa}^{[2n+1]}_1}$. 

\section{BFKL-DGLAP Interpolating Curves}\label{sp:InterpolatingCurves}
\begin{figure}[!ht]
\minipage{0.32\textwidth}
  \includegraphics[width=\linewidth]{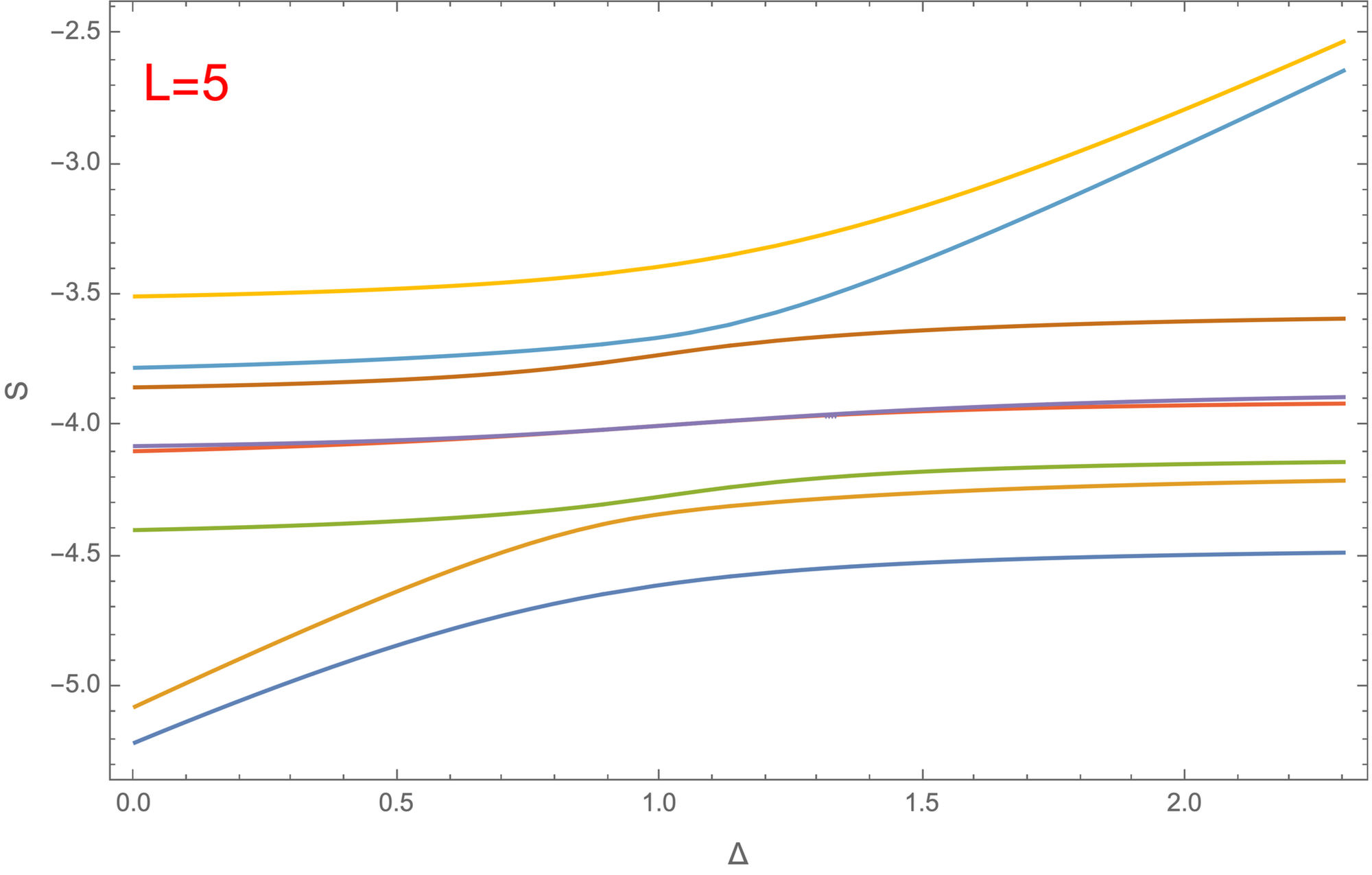}
\endminipage\hfill
\minipage{0.32\textwidth}
  \includegraphics[width=\linewidth]{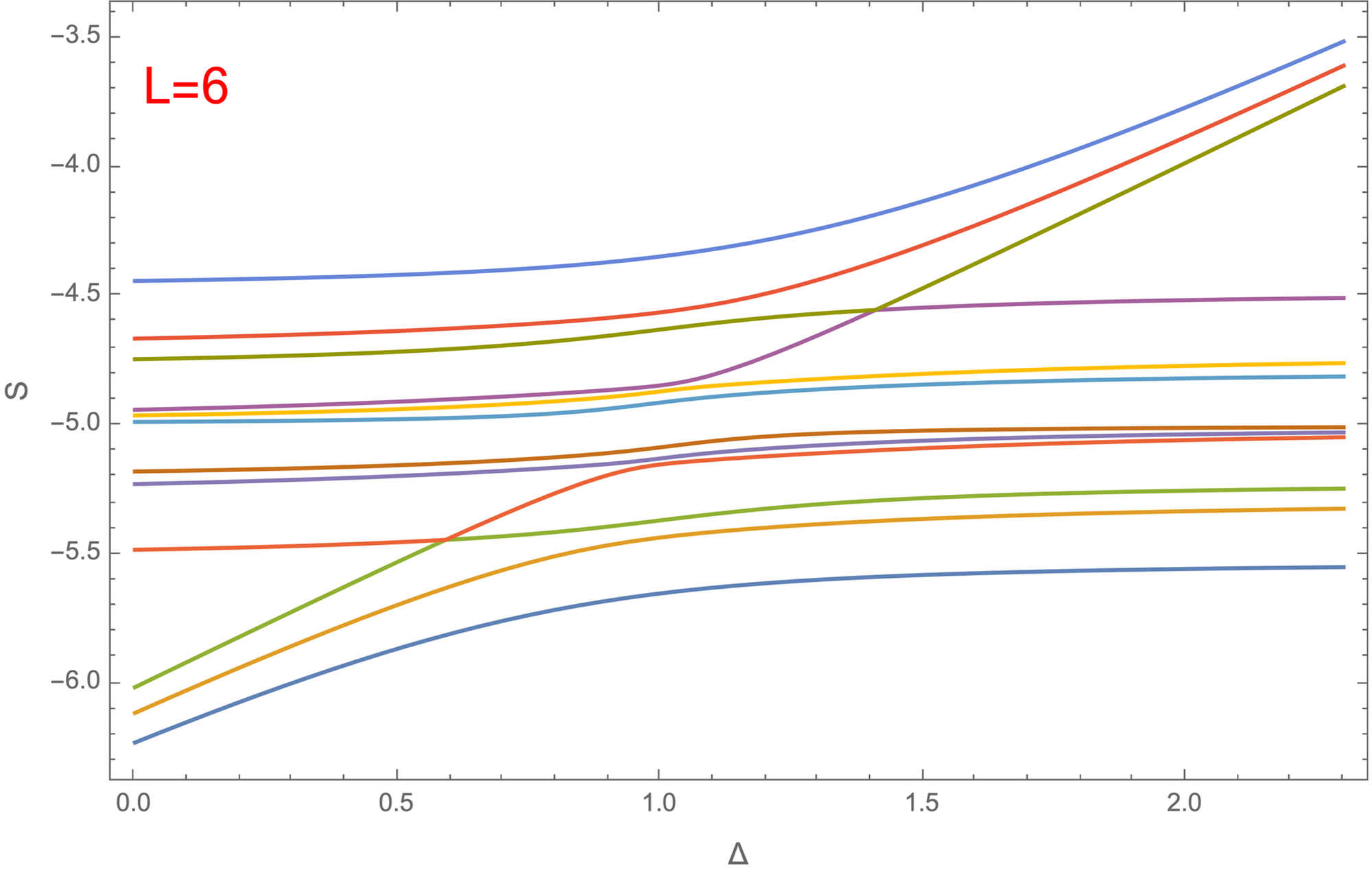}
\endminipage\hfill
\minipage{0.32\textwidth}%
  \includegraphics[width=\linewidth]{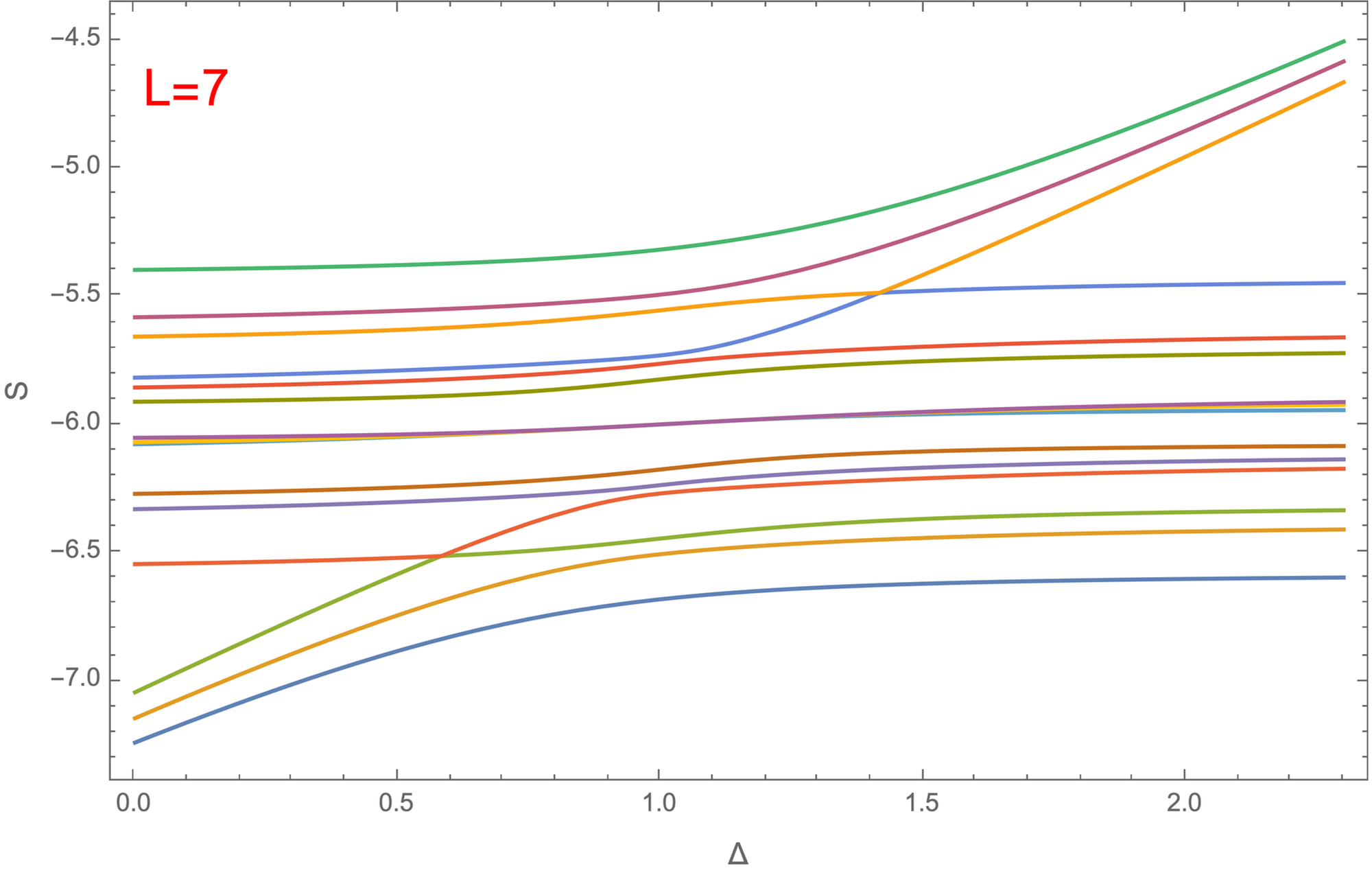}
\endminipage
\caption{Interpolating curves, describing the branching process from HT to DT for $L=5,6,7$ at $g=1/10$.}\label{fig:branchhigherL}
\end{figure}
Here we present the results for the analog of \eqref{curveL3} and \eqref{curveL4} for higher $L$. When going to higher $L$ we get increased amount of data, which constrain the $(\Omega,{\mathbb D})$ interpolating curve. At the same time we also have less numerical data on the number of trajectories mixing from DGLAP side. We made an assumption that in general there are $\left\lfloor \frac{L}{2}\right\rfloor$ diagonal trajectories mixing with the first set of horizontal trajectories. This assumption is verified by a over-determined data from ${L+1}$ orders in $g$ of $\lfloor\frac{L^2}{4}\rfloor$ horizontal trajectories, making us very confident about this assumption, but it would be interesting to verify it directly from $\mathfrak{sl}(2,\mathbb{R})$ light-ray Hamiltonian.

We found the interpolating curves explicitly for $L=3,\dots,11$. The first two are given in \eqref{curveL3} and \eqref{curveL4}, the next three are given below and in FIG.~\ref{fig:branchhigherL}. We are happy to provide the remaining curves upon request. For $L=5$ we get:
\beqa\nonumber\la{Q11sol}
\Omega ^8-2 {\mathbb D} \Omega ^7+\left({\mathbb D}^2-56\right) \Omega ^6+80 {\mathbb D} \Omega ^5+\left(784-24 {\mathbb D}^2\right) \Omega ^4-672 {\mathbb D} \Omega ^3+\left(80 {\mathbb D}^2-3136\right) \Omega ^2+896 {\mathbb D} \Omega -64 {\mathbb D}^2\;,
\eeqa
for $L=6$ we get:
\begin{align}\nonumber
&\Omega ^{12}-3 {\mathbb D} \Omega ^{11}+\left(3 {\mathbb D}^2-80\right) \Omega ^{10}+\left(200 {\mathbb D}-{\mathbb D}^3\right) \Omega ^9+\left(2016-160 {\mathbb D}^2\right) \Omega ^8+\left(40 {\mathbb D}^3-4032 {\mathbb D}\right) \Omega ^7\\
\nonumber&+\left(2400 {\mathbb D}^2-19456\right) \Omega ^6+\left(29184 {\mathbb D}-384 {\mathbb D}^3\right) \Omega ^5+\left(64512-11264 {\mathbb D}^2\right) \Omega ^4+\left(1024 {\mathbb D}^3-64512 {\mathbb D}\right) \Omega ^3\\
\nonumber&+\left(12288 {\mathbb D}^2-81920\right) \Omega ^2+40960 {\mathbb D} \Omega +32768\;,
\end{align}
for $L=7$ we get:
\begin{align}
\nonumber&\Omega ^{15}-3 {\mathbb D} \Omega ^{14}+\left(3 {\mathbb D}^2-108\right) \Omega ^{13}+\left(276 {\mathbb D}-{\mathbb D}^3\right) \Omega ^{12}+\left(4176-228 {\mathbb D}^2\right) \Omega ^{11}+\left(60 {\mathbb D}^3-8816 {\mathbb D}\right) \Omega ^{10}\\
\nonumber&+\left(5744 {\mathbb D}^2-73152\right) \Omega ^9+\left(121920 {\mathbb D}-1104 {\mathbb D}^3\right) \Omega ^8+\left(601344-58432 {\mathbb D}^2\right) \Omega ^7\\
\nonumber&+\left(7616 {\mathbb D}^3-734976 {\mathbb D}\right) \Omega ^6+\left(233216 {\mathbb D}^2-2239488\right) \Omega ^5+\left(1741824 {\mathbb D}-17664 {\mathbb D}^3\right) \Omega ^4+\left(2985984-304128 {\mathbb D}^2\right) \Omega ^3\\
\nonumber&+\left(15360 {\mathbb D}^3-995328 {\mathbb D}\right) \Omega ^2+110592 {\mathbb D}^2 \Omega -4096 {\mathbb D}^3\;.
\end{align}
These polynomials are expected to control the leading singularities in $\frac{g^{n}}{(\Delta-1)^{n-1}}$ to all orders in perturbation theory for all HTs
at given $L$ as well the leading $\frac{g^{2n}}{\omega^{2n-1}}$ divergence to all orders in $g$ for all DTs, which branch into HTs on the first instance.

\subsection{Branching spectrum for $L=3,\dots,11$}\label{sp:BranchingSpectrum}
From the interpolating curves in the previous section we can extract all leading divergences in $1/\omega$ to all orders in $g$. Let us focus on one loop divergence which is given by 
$\Delta-1=\omega+\frac{b_n g^2}{\omega}+\dots$
where the {\it branching spectrum} $\{ b_n\}$ is a collection of residues. For a given $L$, $b_n$ can be obtained as roots of a characteristic polynomial, which can be obtained by taking the corresponding limit of the interpolating curves. We found the following polynomials
$$
\begin{array}{c|l}
L&\text{characteristic polynomial}\\ \hline
 3 & b+16 \\
 4 & b^2+24 b+64 \\
 5 & b^2+32 b+192 \\
 6 & b^3+40 b^2+384 b+512 \\
 7 & b^3+48 b^2+640 b+2048 \\
 8 & b^4+56 b^3+960 b^2+5120 b+4096 \\
 9 & b^4+64 b^3+1344 b^2+10240 b+20480 \\
 10 & b^5+72 b^4+1792 b^3+17920 b^2+61440 b+32768 \\
 11 & b^5+80 b^4+2304 b^3+28672 b^2+143360 b+196608 \\
\end{array}
$$
Based on the above one can guess that the coefficients in the polynomial are given by a simple formula $\frac{(-1)^n 2^{3 n} (n-L)_n}{n!}$, which leads to the following general expression for the branching spectrum characteristic polynomial
$$
P_L(b)=b^{\left\lceil \frac{L-1}{2}\right\rceil } \, _2F_1\left(\frac{1}{2}-\frac{L}{2},-\frac{L}{2};-L;-\frac{32}{b}\right)\;.
$$

\subsection{One loop $\mathfrak{sl}(2)$ two magnon solution}
In this subsection we argue that there is an intriguing relation between the branching spectrum, discussed in the previous section, and the one loop anomalous dimension of the local operators. Here we consider the simplest two magnon solution, which is constrained by $u_1=-u_2$ due to the cyclicity  symmetry of the trace. The anomalous dimension can be computed from the $\mathfrak{sl}(2)$ Bethe ansatz which in this case reads as
$$
\left(\frac{u_1+i/2}{u_1-i/2}\right)^L = \frac{u_1-i/2}{u_1+i/2}\;\;,\qquad\gamma^{(1)} = \frac{4}{u_1^2+1/4}\;,
$$
introducing $e^{ip}=\frac{u_1+i/2}{u_1-i/2}$ we get
\beq\la{onlocal}
p_k = \frac{2 n \pi }{L+1}\;\;,\;\;\gamma^{(1)}_n=
16 \sin ^2\left(\frac{n \pi }{L+1}\right)\;\;,\;\;n=1,\dots,\left\lceil \frac{L-1}{2}\right\rceil\;.
\eeq
It is convenient to introduce a polynomial $p_L(\gamma)=\prod_n (\gamma-\gamma_n^{(1)})$. We notice that $p_L(16+b/2)\propto P_L(b)$, which establishes a quite non-trivial relation between the spectrum of local operators and the residues of the first singularity in the Regge trajectories. 

\section{Explicit Results for $\omega$}
In this section we give explicit results for $\omega$ as computed from the ABBA. First, let us give the explicit result for $\omega$ up to $g^3$ term. We attached a {\it Mathematica} notebook to expand up to any order (and tested it for $g^{10}$). Below is the explicit result for general state at any $L$ for up to $g^4$, in order to give an idea of the general structure
\begin{equation}\la{moreterms}
\scalebox{0.5}{$\begin{split}
&\omega=2 i \left(z_1+z_2-\frac{1}{z_2}-\frac{1}{z_1}\right) g-\frac{\chi (\Delta ) \left(\frac{4 \left(z_1^4+1\right)}{z_1^2}+\frac{4 \left(z_2^4+1\right)}{z_2^2}+16\right) g^2}{L+2}
+\left(\frac{12 i \chi (\Delta )^2 \left(-z_1^3-z_1-\frac{\left(z_2^2-1\right) \left(z_2^2+1\right){}^2}{z_2^3}+\frac{1}{z_1}+\frac{1}{z_1^3}\right)}{(L+2)^2}-\frac{2 i \pi ^2 \left(z_1+z_2\right) \left(z_1 z_2-1\right) \left(2 z_2^2 z_1^4+\left(z_2-z_2^3\right) z_1^3+2 \left(z_2^2+1\right){}^2 z_1^2+z_2 \left(z_2^2-1\right) z_1+2 z_2^2\right)}{3 (L+2) z_1^3 z_2^3}\right) g^3\\
&+\left(\frac{128 \left(z_1^4+z_1^2+z_2^4+z_2^2+\frac{1}{z_2^2}+\frac{1}{z_2^4}+\frac{1}{z_1^2}+\frac{1}{z_1^4}\right) \chi (\Delta )^3}{3 (L+2)^3}+\frac{8 \left(z_1^4+\left(6 z_2^2+16+\frac{6}{z_2^2}\right) z_1^2+z_2^4+16 z_2^2+\frac{16}{z_2^2}+\frac{1}{z_2^4}+36+\frac{6 z_2^2+16+\frac{6}{z_2^2}}{z_1^2}+\frac{1}{z_1^4}\right) \chi ''(\Delta )}{3 (L+2)}\right.\\
&\left.+\frac{4 \pi ^2 \left(8 z_2^4 z_1^8+3 z_2^3 \left(z_2^2-1\right) z_1^7+2 \left(z_2^6+6 z_2^4+z_2^2\right) z_1^6+3 z_2 \left(z_2^6+2 z_2^4-2 z_2^2-1\right) z_1^5+4 \left(z_2^2+1\right){}^2 \left(2 z_2^4-z_2^2+2\right) z_1^4-3 z_2 \left(z_2^6+2 z_2^4-2 z_2^2-1\right) z_1^3+2 \left(z_2^6+6 z_2^4+z_2^2\right) z_1^2-3 z_2^3 \left(z_2^2-1\right) z_1+8 z_2^4\right) \chi (\Delta )}{3 (L+2)^2 z_1^4 z_2^4}\right.\\
&\left.+\frac{4 \left(2 z_2^4 z_1^8-3 z_2^3 \left(z_2^2-1\right) z_1^7+\left(6 z_2^6+32 z_2^4+6 z_2^2\right) z_1^6-3 z_2 \left(z_2^6+2 z_2^4-2 z_2^2-1\right) z_1^5+2 \left(z_2^8+16 z_2^6+48 z_2^4+16 z_2^2+1\right) z_1^4+3 z_2 \left(z_2^6+2 z_2^4-2 z_2^2-1\right) z_1^3+\left(6 z_2^6+32 z_2^4+6 z_2^2\right) z_1^2+3 z_2^3 \left(z_2^2-1\right) z_1+2 z_2^4\right) \zeta (3)}{3 (L+2) z_1^4 z_2^4}\right) g^4+O\left(g^5\right)
\end{split}$}
\end{equation}
here $z_a=-i e^{\frac{i\pi n_a}{L+2}}$.

\subsection{Twist 4 example and numerics}
As a particular more explicit example let us consider twist $L=4$. In this case wrapping hits at $g^{6}$ and so the expressions obtained from ABBA are valid to $\sim g^{5}$. We use the notation $\omega_{n_1,n_2}$ where $(n_1,n_2)=(1,3),(1,5),(2,4),(3,5)$ as shown in the main text in TABLE \ref{tab:my_label}. We find the following results for the $4$ HTs
\begin{equation}
\begin{split}\label{L4explicit}
\omega_{1,3}(g)=\,&2 \sqrt{3} g-\frac{10}{3} \chi (\Delta )g^2\,- \frac{\left(\chi (\Delta )^2+2 \pi ^2\right)}{\sqrt{3}}g^3+\frac{4}{81} \left(369 \chi ''(\Delta )+8 \chi (\Delta )^3+18 \pi ^2 \chi (\Delta )+531 \zeta_3\right)g^4\\
&\qquad\qquad\qquad\pm\frac{(5 \chi (\Delta ) (9360 \chi ''(\Delta )-\chi (\Delta ) (115 \chi (\Delta )^2-324 \pi ^2)+3312 \zeta_3)+2376 \pi ^4)}{1620 \sqrt{3}} g^5+\mathcal{O}\left(g^6\right)\\
\omega_{1,5}(g)=&-\frac{4}{3}  \chi (\Delta )g^2+\frac{4}{81}  \left(-16 \chi (\Delta )^3-9 \pi ^2 \chi (\Delta )+72 \left(\chi ''(\Delta )+\zeta_3\right)\right)g^4+\mathcal{O}\left(g^6\right)\\
\omega_{2,4}(g)=&-4  \chi (\Delta )g^2+\left(\frac{4}{3} \pi ^2 \chi (\Delta )+32 \left(\chi ''(\Delta )+\zeta_3\right)\right)g^4 +\mathcal{O}\left(g^6\right)\\
\omega_{3,5}(g) =&\;\omega_{1,3}(-g)\;.
\end{split}
\end{equation}
We furthermore verified these expressions at $\Delta=\frac{1}{2}$ and $\Delta=0$ using high-precision numerics.

\subsubsection{The Riemann surface}
\begin{figure}[!t]
\minipage{0.49\textwidth}
  \includegraphics[width=\linewidth]{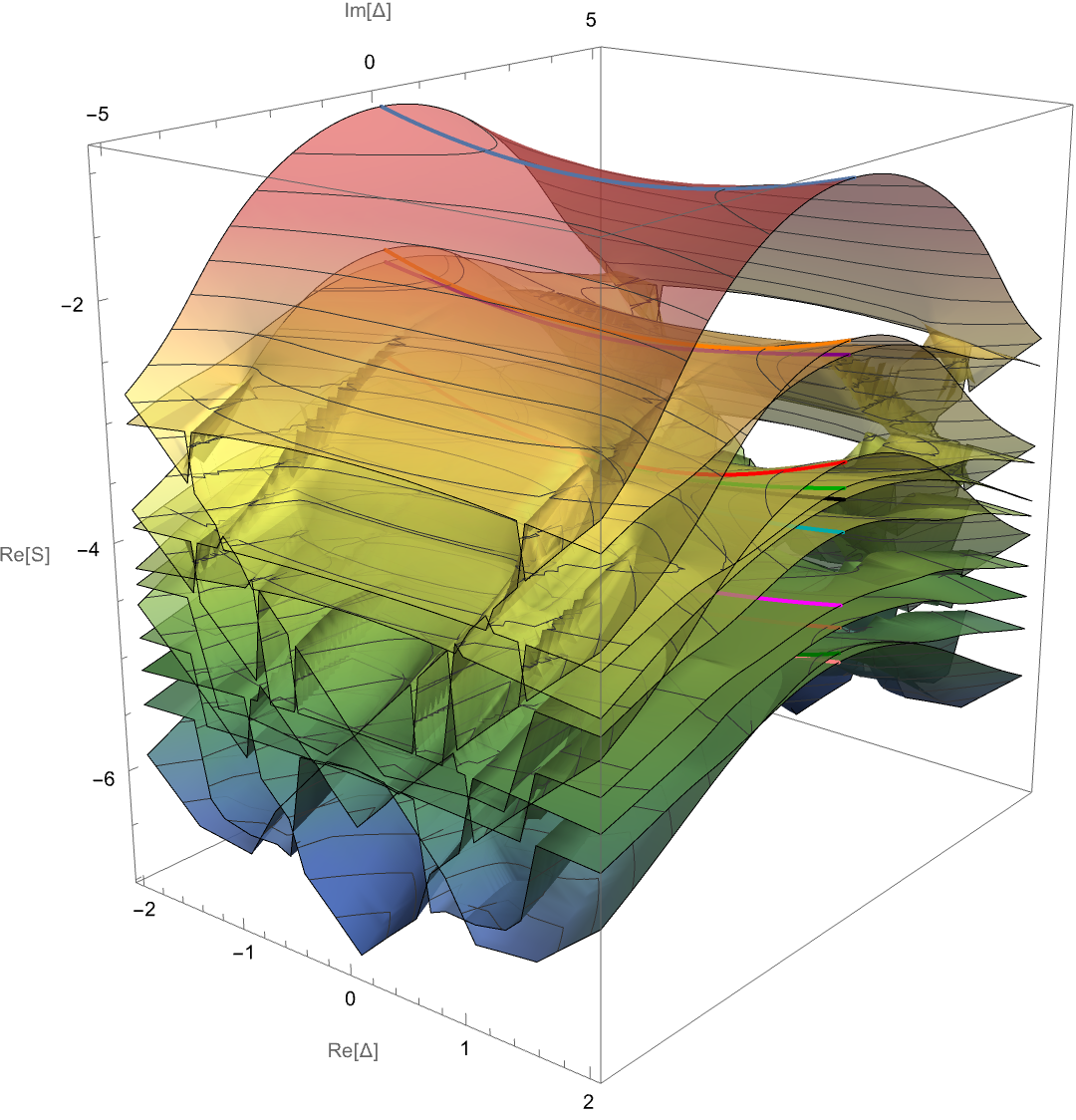}
\endminipage\hfill
\minipage{0.49\textwidth}%
  \includegraphics[width=\linewidth]{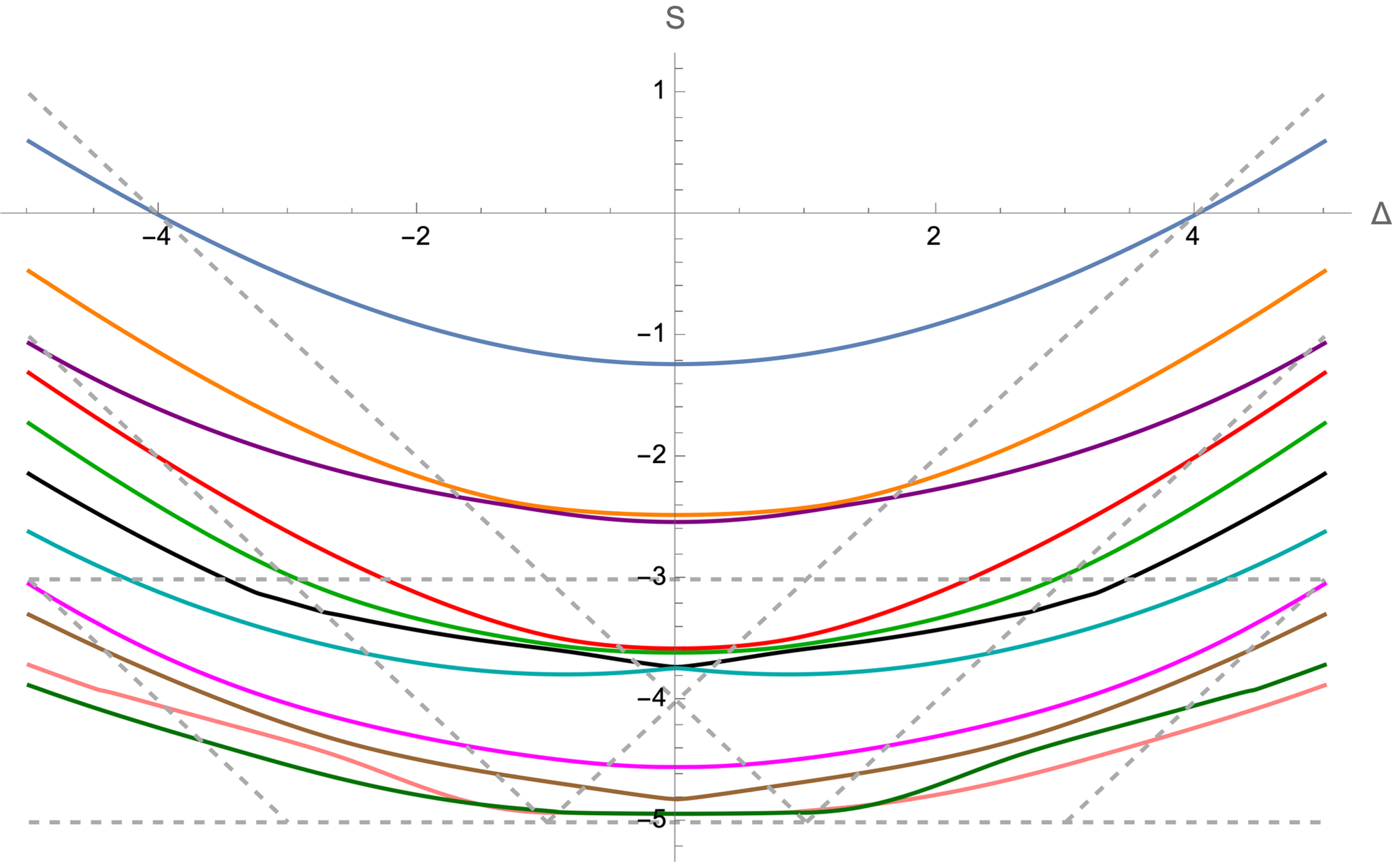}
\endminipage
\caption{On the left: eleven sheets of the Riemann surface connecting
the different trajectories for $L=4$ at $g = 1/2$. The surface is generated with over 38000 points. On the right: the real plane slice of the Riemann surface. The color scheme for trajectories is the same of FIG.~\ref{fig:enter-label1} and FIG.~\ref{fig:branching}. Additional ones are found passing through the various branch points of the Riemann surface.}\label{fig:Riemann}
\end{figure}
 Our numerical strategy is similar to the one  of \cite{Klabbers:2023zdz}. We initialize our numerics with weak coupling data of local operators computed using \cite{Marboe:2017dmb} and then we move along the Regge trajectory varying $\Delta$ to obtain $S$. This lead to all the data for $L=4$ we present in this paper, in particular the trajectories in FIG. \ref{fig:enter-label1}, FIG.~\ref{fig:branching} and FIG.~\ref{fig:Riemann}. We can also fix both $\Delta$ and $S$ while we vary the coupling $g$ moving to the non-perturbative regime. In this case, we can compute the intercepts as presented in FIG.~\ref{fig:intercepts}.

 In order to probe the mixing between the HTs and DTs for $L=4$, we map out the neighborhood of the first HT for complex $\Delta$. For $g=1/2$, we obtain an intricate Riemann surface that presents several quadratic branch points connecting various sheets (see FIG.~\ref{fig:Riemann}). The presence of the branch points allows the trajectories to mix as shown above. When $g$ goes to zero, these branch points approach the real $(\Delta,S)$ plane disconnecting DTs ad HTs at exactly $g=0$.

\subsubsection{The intercepts}
\begin{figure}[t]
    \centering
    \includegraphics[width=.5\textwidth]{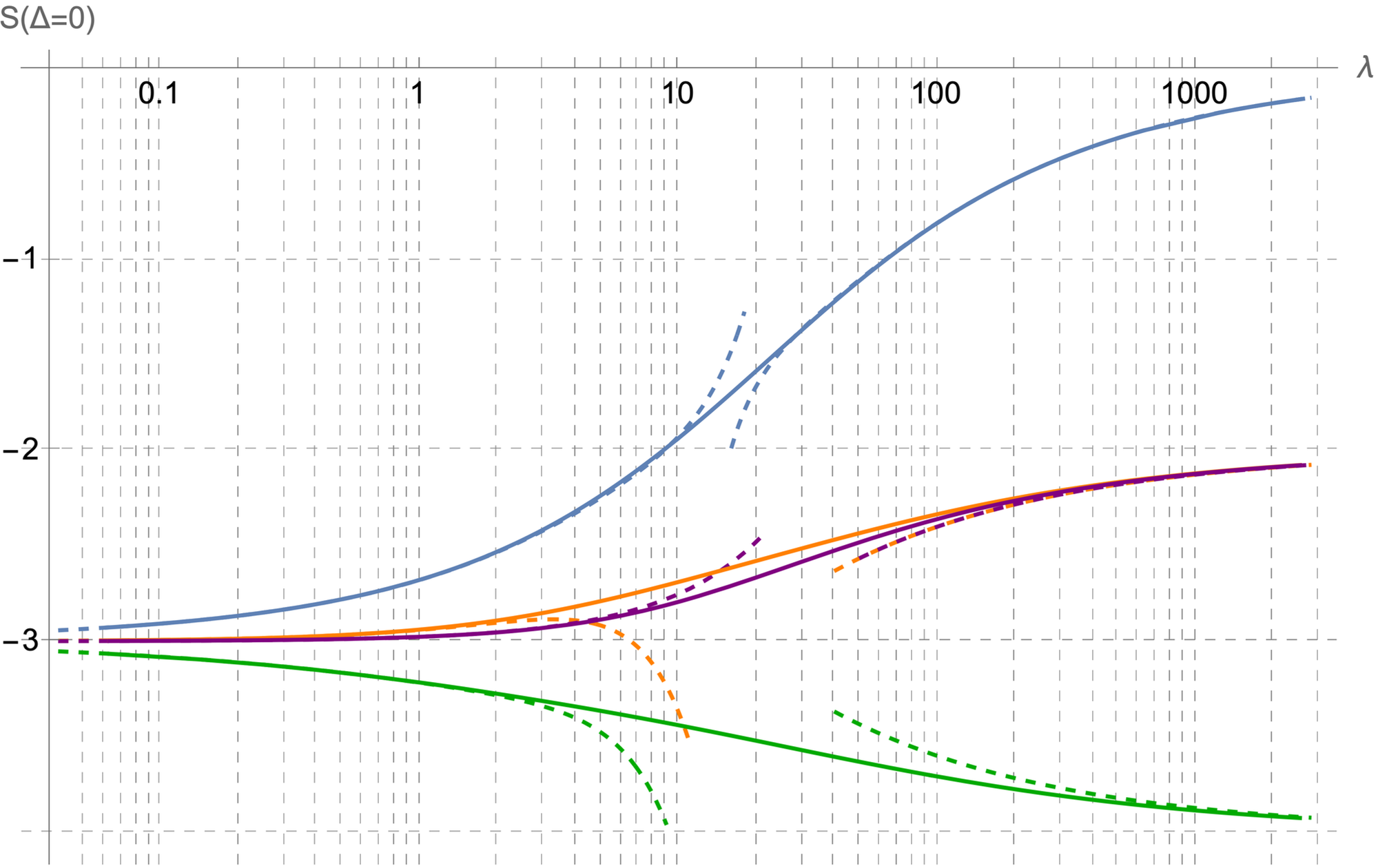}
    \caption{The four intercepts of the $L=4$ horizontal trajectory as function of the coupling $\lambda=16\pi^2 g^2$ with $g=[1/50,4]$. Solid lines are numerical data, dashed lines are the weak and strong coupling prediction/fit. At weak coupling $S(\Delta=0)=-3+\omega(\Delta=0)$, where four solution for $\omega$ are given in \eqref{L4explicit}. At strong coupling, the leading intercept is given in \eqref{intercept1strong}, while the remaining three in \eqref{intercept234strong}. The color scheme is the same of FIG.~\ref{fig:enter-label1} and FIG.~\ref{fig:branching}.}
    \label{fig:intercepts}
\end{figure}
 The value of $\omega$s \eqref{L4explicit} at $\Delta=0$ corresponds to the Regge intercepts. In FIG.\ref{fig:intercepts} we present the non-perturbative numerical data for the four intercepts together with the expected weak coupling behavior given by \eqref{L4explicit} at $\Delta=0$. The strong coupling intercept for the leading trajectory (blue one) was found in \cite{Brower:2014wha,Ekhammar:2024rfj} and it reads
\begin{equation}\label{intercept1strong}
\begin{split}
    S(0) \!=\! -\frac{8}{\sqrt{\lambda }}-\frac{4}{\lambda }+\frac{13}{\lambda ^{3/2}}+\frac{41\!+\!96 \zeta_3\!}{\lambda^2}
    +\frac{\frac{1249}{16}\!+\!288 \zeta_3\!}{\lambda ^{5/2}}
    +\frac{\frac{671}{4}\!+\!192 \zeta_3\!-\!720 \zeta_5\!}{\lambda^{3}}+
    \frac{\frac{71333}{128}\!-\!771\zeta_3\!-\!7170\zeta_5\!-\!4608 \zeta_3^2\!}{\lambda ^{7/2}}\!+...\!   
\end{split} 
\end{equation}
The strong coupling intercept for the remaining three trajectories is unknown. However, fitting our numerical data we obtain the following expansions
\begin{equation}\label{intercept234strong}
S(0)=-2-\frac{4}{\sqrt{\lambda}}\quad\text{(Orange and Purple in FIG.\ref{fig:intercepts})},\qquad\qquad
S(0)=-4+\frac{4}{\sqrt{\lambda}}\quad\text{(Green in FIG.\ref{fig:intercepts})}\;.
\end{equation}

\end{document}